\input phyzzx
\Pubnum={}
\date={}
\pubtype={}
\titlepage
\title{LYAPUNOV EXPONENTS, PATH-INTEGRALS AND FORMS}
\author{E.Gozzi$^{\flat}$ and M.Reuter$^{\sharp}$}
\address{$\flat$ Dipartimento di Fisica Teorica, Universit\`a di Trieste,\break
Strada Costiera 11, P.O.Box 586, Trieste, Italy \break and INFN, Sezione 
di Trieste.\break
\break
$\sharp$ Deutsches Elektronen-Synchrotron DESY,\break Notkestrasse 85, 
D-22603 Hamburg , Germany}
\abstract
In this paper we use a path-integral approach to represent the Lyapunov
exponents of both deterministic and stochastic dynamical systems.
In both cases the relevant correlation functions are obtained
from a (one-dimensional) supersymmetric field theory whose Hamiltonian,
in the deterministic case, coincides with the Lie-derivative of the
associated Hamiltonian flow. The generalized
Lyapunov exponents turn out to be related to the partition functions
of the respective super-Hamiltonian restricted to the spaces of fixed
form-degree.

\vskip 1in
\centerline {Published in: {\bf Chaos, Solitons and Factals, Vol.4 no 7 (1994)
1117}}
\endpage
\chapter{INTRODUCTION}
During the past two decades a lot of efforts have been devoted to the
study of the so-called "stochastic" properties of deterministic
dynamical systems. Most of the work concentrated on finding
appropriate order parameters which could be used to classify
dynamical systems according to their degree of stochasticity\Ref\reic{For a
review see:\nextline L.E.Reichl, "{\it The transition to chaos in conservative
classical systems:\break quantum manifestation}",
Springer-Verlag, New York, 1992}
\Ref\lic{A.J.Lichtenberg and A.M.
Lieberman, {\it "Regular and stochastic motion"},\nextline
Springer-Verlag, New York 1983}.
Rigorous definitions of integrable, KAM-, ergodic, weakly mixing,
mixing, C-systems, etc. were given and it was established that this
ordering of the systems amounts to an increasing "chaoticity" or
"stochasticity" of the motion. It also became clear that some of the
stochasticity properties of the system (like ergodicity) were encoded in the
spectrum of its Liouville operator.
\par
One of the order parameters which has been studied most
extensively is the Kolmogoroff-Sinai (KS) entropy
and similar entropy-like quantities\Ref\casar{M.Casartelli
et al., Phys.Rev.A 13 (1921) 1976;\nextline
G.Benettin et al., Phys.Rev.A 14 (2338) 1976}. A priori,
the KS entropy is defined in a rather abstract manner
\foot{ See ref.[3]~for the
definition and the original references.}. It  has been shown,
\Ref\pes{Ya.B.Piesin, Funcl.Anal.Applic. 8 (1974) 263.} however, that it is
related to the sum of the positive Lyapunov exponents
\Ref\liap{A.M.Lyapunov, {\it "General
problem of stability of motion"},\nextline
Ann.Of Math.Studies no.17, Princeton Univ.Press}.
These exponents, loosly speaking, are a measure of the rate at which
nearby trajectories fly away from each other. For deterministic
systems the Lyapunov exponents are computed from the properties
of a single trajectory, \ie, they are labelled by the initial point of
the respective trajectory.
More recently the concept of Lyapunov exponents has been generalized also
to stochastic systems\Ref\arno{L.Arnold and V.Wihstutz,~"{\it Lyapunov
exponents"}, Lecture Notes in Math. Vol 1186,
Springer-Verlag, New York, 1986}. In this case the
Lyapunov exponents, like all observables, are obtained by
averaging over (infinitely) many trajectories. This averaging can be
done in different, inequivalent ways [see ref.9].
It is well known
\Ref\pari{M.V.Feigelman and A.M.Tsvelik, Sov.Phys.JETP {\bf 56} (1982)
823.;\nextline
G.Parisi and N.Sourlas, Nucl.Phys.{\bf B202} (1982) 321;\nextline
F.Langouche et al.{\it "Functional integration and
semiclassical expansion}",\nextline
Reidel, Dordrecht,  1982}
that stochastic systems with a Langevin dynamics can be formulated via
path-integrals and, once the noise is integrated away, they are
equivalent to a one-dimensional supersymmetric field theory. As was
pointed out before
\Ref\prim{R.Benzi, G.Paladin, G.Parisi and A.Vulpiani, \nextline
J.Phys. A. 18 (1985) 2157}\Ref\copion{R.Graham, Europhys.Lett. 5 (1988) 101}
there exists a special
class of generalized Lyapunov exponents which acquires a
natural interpretation in terms of this supersymmetric field theory.
In fact they are simply given by the lowest eigenvalues of the
super-Hamiltonian at fixed fermion number. Hence
they can be deduced from the asymptotic behaviour of the
corresponding partition function.
\par
The relation between Lyapunov exponents and supersymmetry is much more
general than it might appear from the previous work on stochastic
systems. In fact, let us consider a trajectory~$\phi(t)$~ of some
dynamical system, and let us visualize ~$\phi(t)$~ as a "bosonic
field" in a one-dimensional field theory with t and ~$\phi$~
parametrizing the "spacetime" and the "target-space", respectively.
Let us assume that the lagrangian {\cal L} of the system is invariant
under a supersymmetry transformation of the form
$$\delta\phi(t)=\epsilon c(t)\eqn\benzina$$
where ~$c(t)$~ is the anticommuting "super-partner" of
~$\phi(t)$. Furthermore, assume that {\cal L} leads to an equation
of motion for ~$\phi$~ which is of the general form
$${\dot\phi}(t)={\bf V}(\phi(t))+\eta(t)\eqn\gasoline$$
where ~${\bf V}$~ is some vector field on the "target-
space" and ~$\eta(t)$~ is any function independent of ~$\phi$,
and which does not change under the supersymmetry transformation.
Then supersymmetry implies that the "fermionic partner"
~$c(t)$~ evolves according to
$${\dot c}(t)={\bf V}^{\prime}(\phi(t))c(t)\eqn\diesel$$
where~${\bf V}^{\prime}$~ is the Jacobi matrix (see below)
of ~${\bf V}$. Obviously the equation of motion of the
super-partner ~$c(t)$~ is the linearization of the "bosonic"
equation of motion ~\gasoline. Hence the dynamics of
~$c(t)$~ contains information about the stability
properties of the ~$\phi-$trajectories. In particular,
if two nearby trajectories fly away from each other
exponentially fast, this will manifest itself by
exponentially growing eigenmodes of ~$c(t)$.
\par
The above argument about the connection between supersymmetry and (generalized)
Lyapunov exponents is very
general. It applies to stochastic and deterministic
systems alike. However, only for stochastic systems the field theory
approach we mentioned has been widely used in the literature
\refmark{8,9}, whereas for deterministic systems the relevant
field theory was introduced only recently.
\Ref\goz{E.Gozzi, M.Reuter and W.D.Thacker, Phys.Rev.D 40 (1989) 3363;\nextline
E.Gozzi, M.Reuter, Phys.Lett.233B (1989) 383;E.238B (1990) 451;\nextline
E.Gozzi, M.Reuter, Phys.Lett. 240B (1990) 137;\nextline
E.Gozzi, M.Reuter, W.D.Thacker,~Phys.Rev.D46 (1992) 757;\nextline
E.Gozzi, M.Reuter, W.D.Thacker, Chaos, Solitons and Fractals \break
Vol.2 (1992) 441}
In the stocastic case the pertinent field theory is the euclidean-time
version of supersymmetric quantum mechanics (see ref.[16] for
details). In this context the equation of motion ~$\gasoline$~ is the
Langevin equation with a white noise ~$\eta(t)$. From the point of
view of supersymmetric field theory, eq.~\gasoline~ is the
Nicolai mapping (see the third of ref.[16]) relating ~$\phi$~to the
Gaussian field~$\eta$. In the deterministic case ~$\eta(t)=0$, so that
~\gasoline~ becomes a generic first order evolution equation. We are
particularly interested in Hamiltonian systems, in which case
the vector field ~${\bf V}$~ is the symplectic gradient of some
Hamiltonian H. In ref.[10] we have set up a field theoretic formalism
for such systems, in particular we introduced a
path-integral formulation for deterministic systems. The relevant
Lagrangian is indeed invariant under a supersymmetry of the form
~\benzina.
\par
The paper is organized as follows. In section 2 we briefly review some
basic facts about Lyapunov exponents. Then, in section 3, we introduce
the path-integral for classical hamiltonian systems. In section 4 we
present the observables of this theory which measure the (ordinary)
Lyapunov exponents refering to a single trajectory. Next we show
, in section 5, that the partition function of the classical
super-hamiltonian is related to a set of generalized Lyapunov
exponents which constitute a classical analog of the one found by
Graham\refmark{9} in the stochastic case.
In appendix A we give the corresponding discussion for stochastic
systems. There we generalize previous work for one dimensional
configuration space\refmark{8,9} to higher dimensions in order to
elucidate the geometrical structure underlying the higher
dimensional Lyapunov exponents. In a second appendix (B) instead
we briefly indicate the relation between our path-integral and the
thermodynamic formalism of Ruelle\refmark{23} with the hope to
come back in the future to a more complete study of the relations
between the two.

\chapter{BRIEF REVIEW ABOUT LYAPUNOV EXPONENTS}
Let us consider the differential equation
$${d\over dt}\phi^{a}(t)=h^{a}(\phi(t))\eqn\napoli$$
where ~$h$~ is a vector field on some N-dimensional
manifold~${\cal M}_{N}$~ with local coordinates
~$\phi^{a}$. (Later on ~$h$~ will become the hamiltonian vector
field and \break ${\cal M}_{N}={\cal M}_{2n}$~ a symplectic manifold.)
Let~$\Phi_{cl}^{a}(t;\phi_{0})$~ (the subscript "$\vert_{cl}$"
is for {\it classical}) be the solution of ~\napoli~
with initial condition ~$\Phi_{cl}^{a}(t=0;\phi_{0})=\phi_{0}^{a}$.
We can then define a matrix associated to ~$\Phi_{cl}$~
$$S^{a}_{b}(t;\phi_{0})\equiv{\partial\over\partial\phi_{0}^{b}}
\Phi^{a}_{cl}(t;\phi_{0})\eqn\benevento$$
This matrix is usually known as "Jacobi matrix"\refmark{10}
and it describes how small changes of the initial point affect
the solution at time t. The Jacobi matrix is a solution of the
linear equation
$$\bigl[\partial_{t}\delta_{b}^{a}-\partial_{b}h^{a}(\Phi_{cl}(t;\phi_{0}))
\bigr]S^{b}_{c}(t;\phi_{0})=0\eqn\benevento$$
with ~$S^{a}_{b}(0;\phi_{0})=\delta_{b}^{a}$. Eq.~\benevento~ is also
called the "equation of the first variations", in fact if we displace the
initial point ~$\phi_{0}$~ by an infinitesimal amount
~$\delta\phi^{a}(0)$~ then the dispacement at later
times is given , to first order, by the Jacobi field
$$\delta\phi^{a}(t)=S^{a}_{b}(t;\phi_{0})\delta\phi^{b}(0)\eqn\macerata$$
\par
Let us pick some smooth Riemannian metric on ~${\cal M}_{N}$~
and let ~$e^{a}$~ denote a unit vector in the tangent
space ~$T_{\phi}{\cal M}$.  Then
~$\lambda(t;\phi,e)=\Vert S(t;\phi)e\Vert$~ is called the coefficient
of expansion in the direction of ~$e$~. If
$$\limsup_{t\rightarrow\infty}{1\over
t}ln~\lambda(t;\phi,e)>0\eqn\positano$$
there is then  an exponential divergence in the direction of ~$e$.
In ref.[5] ( and  summarized in the second of ref.[3])~it was shown, under
 rather
general conditions, that:\nextline
\item{ i)} the one dimensional Lyapunov exponent for the direction
of ~$e$, as defined by
$$\lambda^{(1)}(e;\phi)=\limsup_{t\rightarrow\infty}{1\over t}ln\Vert
S(t;\phi)e\Vert\eqn\cosenza$$
exists for every vector ~$e\in T_{\phi}{\cal M}$.
\item{ii)} there exists a basis (called "normal
basis"\refmark{5})~$\{e_{i},1\le i\le N\}$~
for
~$T_{\phi}{\cal M}_{N}$~ \break such that
$$\sum_{i=1}^{N}\lambda^{(1)}(e_{i},\phi)=inf \sum_{i}^{N}
\lambda^{(1)}({\tilde e}_{i},\phi)\eqn\catanzaro$$
where the "infimum" is taken over all possible bases
~$\{{\tilde e}_{i}\}$ of ~$T_{\phi}{\cal M}_{N}$.
It is then easy to realize that for any
$e\in T_{\phi}{\cal M}$, we  have
$$\lambda^{(1)}(e;\phi)\in\{\lambda^{(1)}(e_{i},
\phi); 1\le i \le N\}$$
The numbers ~$\lambda^{(1)}(e_{i};\phi),1\le i \le N$, are called the
one-dimensional Lyapunov characteristic numbers
or exponents  at $\phi$. We shall write\break
~$\lambda_{i}^{(1)}(\phi)\equiv \lambda^{(1)}(e_{i},\phi)$~ and
choose the labelling such that
$$\lambda^{(1)}_{1}(\phi)\ge\lambda^{(1)}_{2}(\phi)\ge\cdots\ge
\lambda^{(1)}_{N}(\phi)\eqn\aquileia$$
The numbers~$\{\lambda^{(1)}_{i}\vert 1\le i \le N\}$~ are not
necessarily distinct.

The higher dimensional Lyapunov exponents ~$\lambda^{(p)}$~ are
defined in a  manner similar to the 1-dimensional case. Let
~$\{e_{1},e_{2},\cdots, e_{p}\}$~ be a set of ~$p\le N$~
orthonormal vectors in ~$T_{\phi}\cal M$. They span a p-dimensional
parallelotope~ $V_{p}\equiv e_{1}\wedge e_{2}\wedge\cdots\wedge e_{p}$.
The exponents ~$\lambda^{(p)}$~ are a measure for the exponential
growth of the volume of $V_{p}$:
$$\lambda^{(p)}(V_{p};\phi)=\limsup_{t\rightarrow\infty}
{1\over t}ln\Vert S(t;\phi)e_{1}\wedge S(t;\phi)e_{2}\wedge\cdots
\wedge S(t;\phi)e_{p}\Vert\eqn\palermo$$
It was shown\refmark{5}
that, for almost all initial ~$V_{p}$'s,
~$\lambda^{(p)}=\lambda_{1}^{(p)}$ where ~$\lambda_{1}^{(p)}$~ is
given by the sum of the p largest one-dimensional Lyapunov exponents
$$\lambda^{(p)}_{1}(\phi)\equiv \sum _{i=1}^{p}\lambda_{i}^{(1)}
(\phi)\eqn\cavo$$
Besides the numbers ~$\{\lambda_{i}^{(1)}(\phi)\}$, which are
uniquely associated to the flow ~$h^{a}$ and the point ~$\phi$,
there is a quantity that is associated to the flow itself
and not to any  point in particular. It is the Kolmogoroff-Sinai (KS) entropy
which we mentioned at the begining and which has played a leading
role in detecting the transition from ordered to stochastic  motion
\refmark{3}. According to a theorem by
Piesin\refmark{4}, the KS entropy can be related to the positive
Lyapunov exponents:
$$KS=\int_{{\cal
M}_{N}}d\phi\bigl[\sum_{\lambda_{i}(\phi)>0}\lambda_{i}(\phi)~\bigr]$$
This is the main reason why the central objects to study are the
Lyaupunov exponents.
\par
So far we have discussed Lyapunov exponents for deterministic systems
only. For stochastic systems various inequivalent versions of
Lyapunov-like quantitites have been studied in the
literature\refmark{6}. Here we only mention the one employed by
Benzi et al.\refmark{8} and by Graham\refmark{9}. For the system
~\benzina~ with ~$\eta$~ a white noise, say,
it is not possible to define Lyapunov exponents for individual
trajectories, but only for averages. In ref.[8,9] a generalized
Lyapunov exponent was defined by
$$\Lambda_{1}=\limsup_{t\rightarrow\infty}{1\over t}ln\VEV{tr
S(t)}\eqn\rosita$$
where ~$\VEV{\cdot}$~ denotes the stochastic average over closed
trajectories of length t. Here ~$S(t)$~ is the Jacobi matrix evaluated
at the end-point of the trajectory (actually in ref.[8,9] only systems
with one degree of freedom where considered so that ~$S(t)$ did not
have indices).

\chapter{THE PATH-INTEGRAL FORMULATION OF  CLASSICAL HAMILTONIAN
DYNAMICS}

In classical mechanics (CM)
the propagator ~$P\bigl(\phi_{2},t_{2}\vert \phi_{1},t_{1}\bigr)$, which gives
 the classical
probability for a particle to be at the point
~$\phi_{2}$~at time ~$t_{2}$, given that it was at the point~$\phi_{1}$~at
time $t_{1}$, is just a delta function
$$P\bigl(\phi_{2},t_{2}\vert
\phi_{1},t_{1}\bigr)={\bf\delta}^{2n}\bigl(\phi_{2}-\Phi_{cl}(t_{2},
\phi_{1})\bigr)\eqn\matra$$
where ~$\Phi_{cl}(t,\phi_{0})$~ is a solution of Hamilton's equation
$${\dot\phi}^{a}(t)=\omega^{ab}\partial_{b}H(\phi(t)),~ with~
\omega^{ab}\omega_{bc}=\delta^{a}_{c}\eqn\simca$$
subject to the initial conditions ~$\phi^{a}(t_{1})=\phi_{1}^{a}$~
Here ~$H$~ is the conventional hamiltonian of a dynamical system
defined on some phase-space ~${\cal M}_{2n}$ with local coordinates
~$\phi^{a},a=1\cdots 2n$~ and a constant symplectic structure~$\omega={1\over
2}\omega_{ab}d\phi^{a}\wedge d\phi^{b}$.
\par
The delta function in ~\matra~ can be rewritten as
$$\delta^{2n}\bigl(\phi_{2}-\phi_{cl}(t_{2},\phi_{1})\bigr)=\Bigl\{\prod_{i=1}^{
 N-1}
\int d\phi_{(i)}
 \delta^{2n}\bigl(\phi_{(i)}-\phi_{cl}(t_{i},\phi_{0})\bigr)\Bigr\}
\delta^{2n}\bigl(\phi_{2}-\phi_{cl}(t_{2},\phi_{1})\bigr)\eqn\isotta$$
where we have sliced the interval  [0,t] in N intervals and
labelled the various instants as ~$t_{i}$ and the fields at ~$t_{i}$~
as
~$\phi_{(i)}$. Each delta function
contained in the product on the RHS of ~\isotta~ can be written as:
$$\delta^{2n}\bigl(\phi_{(i)}-\phi_{cl}(t_{i},\phi_{0})\bigr)=\prod_{a=1}^{2n}
\delta\bigl({\dot\phi}^{a}-\omega^{ab}\partial_{b}H\bigr)_{\vert
t_{i}}
det\bigl[\delta^{a}_{b}\partial_{t}-\partial_{b}\bigl(\omega_{ac}
(\phi)\partial_{c}H(\phi)\bigr)\bigr]_{\vert t_{i}}\eqn\fraschini$$
where the argument of the determinant is obtained from the functional
derivative of the equation of motion~\simca~ with respect
to~$\phi_{(i)}$. Introducing anticommuting variables ~$c^{a}$~and~
${\bar c}_{a}$~ to exponentiate the determinant, and the commuting
auxiliary variables ~$\lambda_{a}$~ to exponentiate the delta
functions, we can re-write the propagator as a path-integral.
$$P\bigl(\phi_{2},t_{2}\vert
\phi_{1},t_{1}\bigr)=\int_{\phi_{1}}^{\phi_{2}}
{\cal D}\phi~{\cal D}\lambda~
{\cal D}c~{\cal D}{\bar c}~exp~i{\widetilde S}\eqn\jaguar$$
where~
${\widetilde S}=\int^{t_{2}}_{t_{1}}dt~{\widetilde{\cal L}}$
with
$${\widetilde{\cal L}}\equiv
\lambda_{a}\bigl[{\dot\phi}^{a}-\omega^{ab}\partial_{b}H(\phi)\bigr]+
i{\bar c}_{a}\bigl(\delta^{a}_{b}\partial_{t}-\partial_{b}
[\omega^{ac}\partial_{c}H(\phi)]\bigr)c^{b}\eqn\roice$$
In the path-integral~\jaguar~ we have used the using the slicing~\isotta
~and then taken the limit  of ~$N\rightarrow\infty$. This limit
has to be taken with some care and some normalization factors
might appear in eq.[3.5], but they are of no importance for
our discussion.
Holding~$\phi$~ and ~$c$~ both fixed at the endpoints of the
path-integral, we obtain the kernel\refmark{10},~$K(\phi_{2},
c_{2},t_{2}\vert
\phi_{1},c_{1},t_{1})$, which propagates distributions in the space
~$(\phi,c)$
$${\widetilde\varrho}(\phi_{2},c_{2},t_{2})=\int d^{2n}\phi_{1}~
d^{2n}c_{1}~{\bf K}(\phi_{2},c_{2},t_{2}\vert \phi_{1},c_{1},t_{1})
{\widetilde \varrho}(\phi_{1},c_{1},t_{1})\eqn\audi$$
The distributions ~${\widetilde\varrho}(\phi,c)$~ are finite sums
of monomials of the type
$${\widetilde \varrho}(\phi,c)={1\over p!}{\varrho}^{(p)}_{a_{1}\cdots
a_{p}}(\phi)~c^{a_{1}}\cdots c^{a_{p}}\eqn\mann$$
The kernel ~${\bf K}(\cdot\vert\cdot)$ is represented by the
path-integral
$${\bf K}(\phi_{2},c_{2},t_{2}\vert \phi_{1},c_{1},t_{1})=\int
{\cal D}\phi^{a}{\cal D}\lambda_{a}{\cal D} c^{a}
{\cal D}{\bar c}_{a}~exp~i\int_{t_{1}}^{t_{2}}dt
{\widetilde {\cal L}}\eqn\ford$$
with the boundary conditions ~$\phi^{a}(t_{1,2})=\phi^{a}_{1,2}$
and ~$c^{a}(t_{1,2})=c^{a}_{1,2}$. The function~
${\widetilde\varrho}$~ of eq.~\mann~ is the {\it classical} analogue
of a wave function in the Schroedinger picture.
It is also easy from here to build a classical generating functional
~$Z_{cl}$~ from which all correlation-functions can be derived.
It is given by
$$Z_{cl}=\int {\cal D}\phi^{a}(t)~{\cal D}\lambda_{a}(t)~{\cal D}
c^{a}(t)~{\cal D}{\bar c}_{a}~exp~i\int dt \bigl\{{\widetilde{\cal L}}+
source~terms\bigr\}\eqn\fiat$$
where the lagrangian can be written as
$${\widetilde{\cal L}}=\lambda_{a}{\dot{\phi}}^{a}+i{\bar c}_{a}{\dot c}
^{b}-{\widetilde{\cal H}}\eqn\lancia$$
with the "superhamiltonian" given by
$${\widetilde{\cal H}}=\lambda_{a}h^{a}+i{\bar c}_{a}\partial_{b}h^{a}
c^{b}\eqn\bmw$$
and where ~$h^{a}$ is the  hamiltonian vector field
$$h^{a}(\phi)\equiv\omega^{ab}\partial_{b}H(\phi)\eqn\maserati$$
From the path-integral ~\fiat~ and ~\lancia~ we can
see\refmark{10} that
the variables~$(\phi,\lambda)$~\break and ~$(c,{\bar c})$~ form
conjugate pairs satisfying the ($Z_{2}$-graded) commutation
relations
$$\eqalign{\bigl[\phi^{a},\lambda_{b}\bigr]\ = & \ i\delta^{a}_{b}\cr
\bigl[c^{a},{\bar c}_{b}\bigr]\ = & \ \delta_{b}^{a}\cr}\eqn\merced$$
The commutators above are defined in precise terms in ref.[10].
Because of these commutators,
in the "Schroedinger-like" picture\refmark{10}, the variables
~$\lambda_{a}$~ and ~${\bar c}_{a}$~ are represented by
~$\lambda_{a}=-i{\partial\over\partial\phi^{a}}\equiv
-i\partial_{a}$~ and by ~${\bar c}_{a}={\partial\over\partial
c^{a}}$. In this way the hamiltonian ~\bmw~ becomes
$${\widetilde{\cal H}}=-il_{h}\eqn\citroen$$
where
$$l_{h}=h^{a}\partial_{a}+c^{b}(\partial_{b} h^{a}){\partial
\over \partial c^{a}}\eqn\renault$$
is the Lie derivative operator. Its
bosonic part coincides with the Liouvillian ~${\hat L}=h^{a}
\partial_{a}$, which gives the evolution of standard distributions
~$\varrho^{(0)}(\phi)$~ in phase-space:
$$\partial_{t}\varrho^{(0)}(\phi,t)=-l_{h}\varrho^{(0)}(\phi,t)=
{\widehat L}\varrho^{(0)}(\phi,t)\eqn\toledo$$
We see from here that our path-integral is nothing else than
the path-integral
counterpart of the operator approach to CM pioneered by
Koopman and von Neumann\Ref\Neum{B.O.Koopman, Proc.Nat.Acad.Sc. USA,
17 (1931) 315;\nextline
J.von Neumann, Ann.Math. 33, (1932) 587}.
The next step is to understand the meaning of the ghosts
~$c^{a}$. From the lagrangian~${\widetilde{\cal L}}$~of~\lancia~
we see that they obey the eq.
$${\dot c}^{a}(t)=\partial_{b}h^{a}(\phi(t))c^{b}(t)\eqn\gaetano$$
which is the same equation as the one for the first
variations~$\delta{\phi}^{a}$~derivable from
eq.~\simca. So we can say that there is a one to one correspondence
between ghosts ~$c^{a}(t)$~ and Jacobi fields ~$\delta\phi^{a}(t)$.
$$c^{a}(t)\sim \delta\phi^{a}(t)\eqn\maria$$
Of course we know that the Jacobi field depends on the trajectory
~$\phi(t)$~ we are varying , but the same is with the
ghosts~$c^{a}$ because in solving~\gaetano~ we had to specify the
classical trajectory~$\phi(t)$~ to insert in ~$h^{a}$.
In a Schroedinger-like picture, in which the ghosts ~$c^{a}(t)$~ do
not depend  on t, but are only arguments of the
~${\widetilde{\varrho}}(\phi,c,t)$,
the correspondence of the ghosts is not with the Jacobi fields but
with the basis of the cotangent space ~$T^{\star}_{\phi}{\cal M}_{2n}$
that is usually indicated as~$d\phi^{a}$. For the details see
ref.[10].
With this interpretation of the ghosts, it is then easy\refmark{10}
to re-interpret all the Cartan-Calculus on symplectic manifolds and also to
identify ~${\widetilde{\cal H}}$~ with the Lie-derivative of the
hamiltonian flow~$l_{h}$appearing in ~\renault.
Because the
ghosts ~$c^{a}$~form a basis of the cotangent space
~$T^{\star}_{\phi}{\cal M}_{2n}$~,~${\widetilde\varrho}(\phi,c)$~ may be
considered a p-form valued field on ~${\cal M}_{2n}$.
The Lie-derivative acts then on the components of
~${\widetilde\varrho}$~\mann~ in the standard
manner
$$l_{h}\varrho^{(p)}_{a_{1}a_{2}\cdots a_{p}}=h^{b}\partial_{b}
\varrho^{(p)}_{a_{1}\cdots a_{p}}+\sum _{j=1}^{p}
\partial_{a_{j}}h^{b}\varrho^{(p)}_{a_{1}\cdots
a_{j-1}ba_{j+1}\cdots a_{p}}\eqn\peugeout$$

The kernel ~$K$~ obeys the Schroedinger-like equation
~$i\partial_{t}K={\tilde {\it H}}K$. Consequently the time
evolution of ~${\tilde\varrho}$~ is governed by the equation
$$i\partial_{t}{\tilde\varrho}(\phi,c,t)={\widetilde{\cal H}}{\tilde
\varrho}(\phi,c,t)\eqn\seat$$
or
$$\partial_{t}\varrho^{(p)}_{a_{1}\cdots a_{p}}(\phi,t)=-l_{h}
\varrho^{(p)}_{a_{1}\cdots a_{p}}(\phi,t)\eqn\ibiza$$

\par
The explicit evaluation\refmark{10}\Ref\Barn{T.Barnes and G.I.Ghandour,
Nucl.Phys. B146 (1978) 483;
\nextline
R.J.Rivers,"{\it Path-integral methods in quantum field
theory}"\nextline
Cambridge University Press, Cambridge,  1987}
of the path-integral
~\ford~ yields
$${\bf K}\bigr(\phi_{2},c_{2},t_{2}\vert \phi_{1},c_{1},t_{1}\bigl)=
{\bf\delta}^{(2n)}\bigl(\phi^{a}_{2}-\Phi^{a}_{cl}(t_{2};\phi_{1})\bigr)
{\bf\delta}^{(2n)}\bigl(c^{a}_{2}-{\bf C}^{a}_{cl}(t_{2};c_{1},
\phi_{1})\bigr)\eqn\malaga$$
Here ~$\Phi^{a}_{cl}(t)$~ and ~${\bf C}^{a}_{cl}(t)$~ are solutions of the
classical equations of motion resulting from ~${\widetilde{\cal L}}$,
$${\dot\phi}^{a}(t)=h^{a}(\phi(t))\equiv
\omega^{ab}\partial_{b}H(\phi(t))\eqn\madrid$$
$${\dot c}^{a}(t)=\partial_{b}h^{a}(\phi(t))c^{b}(t)\eqn\zara$$
with the boundary
conditions~$\Phi^{a}_{cl}(t_{1};\phi_{1})=\phi^{a}_{1}$ and
~${\bf C}^{a}_{cl}(t_{1};c_{1},\phi_{1})=c^{a}_{1}$.\break
The third argument of ~$C^{a}_{cl}(t;c_{1},\phi_{1})$
indicates that ~${\bf C}^{a}_{cl}$~ is the Jacobi field for the
classical trajectory ~$\Phi^{a}_{cl}(t)$~ emanating from the initial
point ~$\phi_{1}$.
Eq.~\zara~ is solved by
$$c^{a}(t)=S^{a}_{b}(t)c^{b}(t_{1})\eqn\siviglia$$
if the Jacobi matrix obeys the differential equation
$$\bigl[\partial_{t}\delta^{a}_{b}-M^{a}_{b}(t)\bigr]S^{b}_{c}(t)
=0\eqn\cordova$$
with the initial condition ~$S^{a}_{b}(t_{1})=\delta^{a}_{b}$
and where
$$ M^{a}_{b}(t)\equiv \partial_{b}h^{a}(\phi(t))\equiv \omega^{ac}
\partial_{b}\partial_{c}H(\phi(t))\eqn\baleari$$
The formal solution to eq.~\cordova~ reads
$$S(t)={\widehat T}exp\int^{t}_{t_{1}}dt^{\prime}~
M(t^{\prime})\eqn\barcellona$$
where ~${\widehat T}$~ denotes the time-ordering operator.
Note that ~$S(t)$~ functionally depends on the trajectory
~$\phi(t)$~ chosen. Since the latter is uniquely characterized by its
initial point~$\phi_{1}$~ we shall write ~$S(t)\equiv S(t;\phi_{1})$~
for the Jacobi matrix of the trajectory emanating from ~$\phi_{1}$.
The function ~$S(t;\phi_{1})$~ defines what is
called\Ref\ose {V.I.Oseledec, Trans.Moscow Math.Soc. 19 (1968) 197} a
"multiplicative cocycle":
$$S^{a}_{b}(t+\tau;\phi_{0})=S^{a}_{c}(t;\Phi_{cl}(\tau;\phi_{0}))
S^{c}_{b}(\tau;\phi_{0})\eqn\valencia$$
It is well known\Ref\arno{V.Arnold, {\it "Mathematical Methods
in Classical Mechanics"}\nextline Springer-Verlag, New York, 1978}
that ~$S$~ is  symplectic,
\foot{This is most easily seen by noting that ~$M$~ is of the
form ~$\omega^{ab}$~ times a symmetric matrix
\Ref\litt{R.G.Littlejohn, Phys.Rep. 138 (1986) 193} and therefore
lies in the lie-algebra of Sp(2n).},
$S \in Sp(2n)$, \ie,
$$S^{a}_{c}\omega_{ab}S^{b}_{d}=\omega_{cd}$$
As a consequence we have that ~$det S=1$~.
Using ~\siviglia~ we write for the Kernel~\malaga~
$${\bf K }(\phi_{2},c_{2},t_{2}\vert \phi_{1},c_{1},t_{1})=
{\bf \delta}^{(2n)}(\phi_{2}^{a}-\Phi_{cl}^{a}(t_{2};\phi_{1}))
{\bf \delta}^{(2n)}(c_{2}^{a}-S^{a}_{b}(t_{2};\phi_{1})
c_{1}^{b})\eqn\bastia$$
In what follows we shall frequently exploit the fact that ~${\bf K }$
is normalized,
$$\int~d^{2n}\phi_{1}~d^{2n}c_{1}~{\bf K }(\phi_{2},c_{2},t_{2}\vert
\phi_{1},c_{1},t_{1})=1\eqn\versailles$$
$$\int~d^{2n}\phi_{2}~d^{2n}c_{2}~{\bf K }(\phi_{2},c_{2},t_{2}\vert
\phi_{1},c_{1},t_{1})=1\eqn\parigi$$
and that it conserves the Grassmannian delta-function
~${\bf \delta}^{2n}(c)$:
$$\int~d^{2n}\phi_{1}d^{2n}c_{1}~{\bf K}(\phi_{2},c_{2},t_{2}\vert
\phi_{1},c_{1},t_{1})~{\bf \delta}^{(2n)}(c_{1})={\bf \delta}
^{(2n)}(c_{2})\eqn\tours$$
In fact, eq. ~\tours~ can be regarded as an expression of Liouville's
theorem\refmark{13}. Recall that the Liouville
measure on phase space is given by
$$d\mu=\omega^{n}=({1\over 2}\omega_{ab}d\phi^{a}\wedge
d\phi^{b})^{n}\eqn\digione$$
Consequently, if ~$\omega_{ab}$~ assumes its canonical,~\ie
,~$\phi$-independent form, we have
$$d\mu=\omega^{n}=n!~d\phi^{1}\wedge d\phi^{2}\wedge\cdots
\wedge d\phi^{2n}\eqn\lione$$
Invoking the correspondence\refmark{10} ~$c^{a}\longleftrightarrow d\phi^{a}$~
between ghosts and differential forms on ~${\cal M}_{2n}$~, we see
that ~$\delta^{2n}(c)$~ corresponds to the Liouville measure
$${\bf \delta}^{(2n)}(c)\equiv c^{1}c^{2}c^{3}\cdots
c^{2n}\longleftrightarrow d\mu\eqn\montpellier$$
Eq.~\tours~ expresses the fact that ~$d\mu$~ is invariant under the
Hamiltonian flow. The infinitesimal version of eq.~\tours~ is
$$l_{h}{\bf \delta}^{(2n)}(c)=0\eqn\infi$$
with the lie-derivative ~$l_{h}$~ defined in ~\renault. Eq.~\infi~
follows from
$$c^{b}{\partial\over\partial
c^{a}}{\bf\delta}^{(2n)}(c)=0\eqn\annemasse$$
and ~$\partial_{a}h^{a}=0$. It implies that not only zero-forms
evolve according to the Liouville equation~\toledo~ but also the
coefficent functions of the 2n-forms
$${\widetilde\varrho}(\phi,c)=\varrho_{1\cdots 2n}(\phi)c^{1}c^{2}\cdots
c^{2n}\equiv \varrho(\phi){\bf\delta}^{(2n)}(c)\eqn\bianco$$
\ie, the equation~$\partial_{t}\varrho=-{\widehat L}\varrho$~
can be immediately
derived from ~\infi. In the following
we shall use the 2n-form~\bianco~ in order to represent conventional
scalar densities on phase-space. From a mathematical point of view the
factor~${\bf\delta}^{(2n)}(c)$~ provides the volume form on
~${\cal M}_{2n}$~,~whereas from a physical point of view
~${\bf\delta}(c)$ is the vacuum of the fermionic
Fock space\foot{See the appendix and ref.[16] for more details
about this.}. In the Schroedinger-like picture mentioned before
\Ref\swed{E.Witten,
Nucl.Phys.  B188 (1981) 513;\nextline P.Salomonson and J.W. Van
Holten, Nucl. Phys. B196  (1982) 509;\nextline
H.Nicolai, Phys.Lett.  89 B (1980)341}
the condition\break ${\hat c}^{a}\ket{vac}=0$~ translates into~$c^{a}<c\vert
vac>=0$ ,
\ie, the "position representation" of 
$\ket{vac}$~ is ~$<c\vert vac>=\delta^{(2n)}(c)$.

To summarize this  section we can say that in the classical
path-integral the ghosts play a double role: a {\it  dynamical} one, in the 
Heisenberg-like\foot{We called it Heisenberg-like
because the variables~$\phi,c$~ depend on t in this "picture".}         
picture of classical mechanics, because their equation of motion is the
Jacobi equation and a {\it geometrical} one , in the Schroedinger-like
picture of CM, because the time-independent ~$c^{a}$~ span the cotangent
space ~$T^{\star}_{\phi}{\cal M}_{2n}$. This double role will
be heavily exploited in our discussion of the Lyapunov
exponents. The dynamical role of the ghosts had
already been partly exploited in the second and the last of ref.[10].
There it was shown that
for any hamiltonian ~$H(\phi)$~ the action~${\widetilde{\cal L}}$~ is
invariant under a set of {\it universal} (graded) symmetries which
form an ISp(2) algebra. Part of this
\endpage algebra consists of a BRS-like
operator ~$Q=ic^{a}\lambda_{a}$~ and an anti-BRS operator
~${\bar Q}=i{\bar c}\omega^{ab}\lambda_{b}$, respectively.
${\widetilde{\cal L}}$~ was also shown to be invariant
(up to surface terms) under
a supersymmetry generated by the charges
$$\eqalign{Q_{H}\ = & \ c^{a}(\partial_{a}-\partial_{b}H)\cr
{\bar Q}_{H} \ = & \ {\bar c}_{a}\omega^{ab}(\partial_{b}+\partial
_{b}H)\cr}\eqn\volvo$$
which are nilpotent, ~$Q_{H}^{2}={\bar Q}_{H}^{2}=0$~ and close on the
superhamiltonian:
$$i{\widetilde{\cal H}}=\bigl[Q_{H},{\bar Q}_{H}\bigr]\eqn\dafca$$
In the second of ref.[10] it was shown that the  phase of a dynamical
hamiltonian system
with this  supersymmetry unbroken  was the same as the ergodic
phase of the system.
Moreover in the last of ref.[10] the supersymmetry and its relation to
ergodicity were applied in a study of the Toda criterion, which was
the first criterion put forward in 1974 to detect transition from ordered
to stochastic motion. The next step , in the physics-history\foot{We
say "physics-history" because in the mathematics literature
 (Kolmogoroff-Sinai ) these tools had been developed before.} of dynamical
systems, was taken in 1974-75\refmark{3} and it consisted
in using more refined tools to study these transitions. These
tools were the KS-entropy and Lyapunov exponents to which we turn now.

\chapter{OBSERVABLES FOR HIGHER DIMENSIONAL LYAPUNOV EXPONENTS}
As we have already stressed
the path-integral formulation\refmark{10} of classical hamiltonian dynamics
naturally involves the Jacobi fields ~$c^{a}(t)$. Therefore it
seems plausible that it should be possible to relate quantities
like the generalized Lyapunov exponents to certain observables
in this theory.

\par
For any observable ~${\cal O}={\cal O}(\phi,\lambda,c,{\bar c})$~
we define the "vacuum expectation value" as
$$\VEV{{\cal O}}=\int {\cal D}\phi {\cal D}\lambda{\cal D} c
{\cal D}{\bar c}~{\cal O}(\phi,\lambda,c, {\bar c})~{\bf \delta}
^{(2n)}(c(-\infty))~exp~i\int_{-\infty}^{\infty}dt
{\widetilde{\cal L}}\eqn\milano$$
where an integration over~$\phi^{a}(\pm\infty)$~ and ~$c^{a}(\pm
\infty)$~ is understood\foot{Here t=$+\infty$~(t=$-\infty$) is a
symbolic notation for some finite time which is larger (smaller) than
any time argument of the fields in ~${\cal O}$.}. Note that here we are
dealing with a {\it trace}-formalism, rather than a {\it bra-ket}
formalism. Therefore, contrary to quantum mechanics, the "state",
a generalized density ~${\widetilde\varrho}$, appears only once under the
path-integral\refmark{10}.
Expectation values of the type \milano~ can be reduced to strings
of propagation kernels~${\bf K}(\phi_{j},c_{j},t_{j}\vert
\phi_{j-1},c_{j-1},t_{j-1})$ connecting field monomials with
different time arguments. An identity which we will often use is
$$\eqalign{\int {\cal D}\phi~{\cal D}\lambda~{\cal D}c~{\cal D}
{\bar c}~\ {\bf A} & \ (\phi(t_{2}),\lambda(t_{2}),c(t_{2}),{\bar c}(t_{2}))
\cdot {\bf B}(\phi(t_{1}),\lambda(t_{1}),c(t_{1}),{\bar c}(t_{1}))\cr
\ \cdot & \ exp\bigl\{i\int_{t_{1}}^{t_{2}}dt
{\widetilde{\cal L}}\bigr\}~{\widetilde\varrho}(\phi(t_{1}), c(t_{1}))=\cr
= \int  \ d\phi_{1} & \ ~dc_{1}~{\bf A}(\phi_{2},-i{\partial\over
\partial\phi_{2}},c_{2},{\partial\over\partial c_{2}})\cr
\ \cdot & \ {\bf K}(\phi_{2},c_{2},t_{2}\vert \phi_{1},c_{1},t_{1})
{\bf B}(\phi_{1},-i{\partial\over\partial\phi_{1}},c_{1},{\partial
\over\partial c_{1}}){\widetilde\varrho}(\phi_{1},c_{1})\cr}\eqn\crema$$
where ~${\bf A}$~ and ~${\bf B}$~ are arbitrary functions and where
the functional integration is subject to the boundary conditions
$\phi(t_{2})=\phi_{2}$ and ~$c(t_{2})=c_{2}$ fixed, while
~$\phi(t_{1})$ and ~$c(t_{1})$ are integrated over.
For the observables considered in the following, ~${\bf A}$
and~${\bf B}$
will be free from ordering ambiguities. Eq.\crema~can be proven by
discretizing the path-integral in the usual way.
\par
To start with
we consider the family of observables
$${\cal O}^{a}_{b}(T;\phi_{0})=c^{a}(T){\bar c}_{b}(0){\bf \delta}
^{(2n)}(\phi(0)-\phi_{0})\eqn\brescia$$
for fixed ~$T>0$~ and ~$\phi_{0}\in {\cal M}_{2n}$. The delta-function
picks the trajectory ~$\phi(t)$~ which passes through the prescribed
point ~$\phi_{0}$ at time t=0. The creation operator
${\bar c}_{b}(0)$~ creates at t=0 a one-ghost state from the vacuum
~${\bf \delta}^{(2n)}(c)$~ and ~$c^{a}(T)$~ destroys it  at some
later time t=T. By using eq.\crema~we obtain for the expectation
value of ~${\cal O}^{a}_{b}$~:
$$\eqalign{\VEV{{\cal O}^{a}_{b}(T, \phi_{0})}\ = & \ \int
d\phi(\infty)~dc(\infty)~{\bf K}(\phi(\infty),c(\infty),\infty\vert
\phi(T),c(T),T)\cr
\ \cdot & \ \int d\phi(T)~dc(T)~c^{a}(T)~\int d\phi(0)dc(0)~{\bf K}
(\phi(T),c(T),T\vert\phi(0),c(0),0)\cr
\ \cdot & \ {\bf \delta}(\phi(0)-\phi_{0})\cr
\ \cdot & \ {\partial\over\partial c^{b}(0)}\int
d\phi(-\infty)dc(-\infty)~{\bf K}(\phi(0),c(0),0\vert \phi(-\infty),
c(-\infty),-\infty)\cr
\ \cdot & \ {\bf\delta}^{(2n)}(c(-\infty))\cr}\eqn\cremona$$
The third factor of ~${\bf K}$~ on the RHS of ~\cremona~ propagates
the vacuum ~${\bf\delta}^{(2n)}(c(-\infty))$\break
from ~$t=-\infty$~to
~$t=0$. Because of Liouville's theorem, eq.\tours, the result at
~$t=0$~ is ~${\bf\delta}^{(2n)}(c(0))$. Then, between t=0 and t=T,
the second factor of ~${\bf K}$~ propagates the ghost excitation created
by ~${\partial\over\partial c^{b}(0)}$~ acting on the delta-function.
For ~$t>T$~ we are left with the vacuum again. Using ~\versailles~ we see
that, again as a consequence of Liouville's theorem, the first ~${\bf K}$~
in ~\cremona~ is ineffective. In this way ~\cremona~
boils down to
$$\eqalign{\VEV{{\cal O}^{a}_{b}(T,\phi_{0})} \ = & \ \int
d\phi(T)~dc(T)~c^{a}(T)~\int dc(0){\bf K}(\phi(T),c(T),T\vert
\phi_{0},c(0),0)\cdot\cr
\ \cdot & \ {\partial\over\partial c^{b}(0)}{\bf\delta}^{(2n)}(c(0))\cr
\ = & \ \int
dc(T)dc(0)~c^{a}(T){\bf\delta}(c(T)-S(T;\phi_{0})c(0))
{\partial\over\partial c^{b}(0)}{\bf \delta}^{(2n)}(c(0))\cr}\eqn\bergamo$$
where also ~\bastia~ has been used . Taking advantage of ~\annemasse~
the final result reads
$$\VEV{{\cal O}^{a}_{b}(T;\phi_{0})}=S^{a}_{b}(T;\phi_{0})
\theta(T)\eqn\mantova$$
The step-function ~$\theta(T)$~ has been included because
~$\VEV{{\cal O}}=0$~ for negative values of T, in  this case
in fact the
destruction operator acts on the vacuum before the creation
operator and therefore the probability is zero. We conclude that the
two-point function of the ghosts is given by the Jacobi matrix
~$S^{a}_{b}$.
\par
Later on we shall see that the higher dimensional
Lyapunov exponents are related
to observables of the form~($0\le f\le 2n$)
$${\cal O}_{f}(T;\phi_{0})=c^{a_{1}}(T)\cdots c^{a_{f}}(T){\bar
c}_{a_{f}}(0)\cdots
{\bar c}_{a_{1}}(0){\bf\delta}(\phi(0)-\phi_{0})\eqn\viadana$$
Their expectation values
$$\Gamma_{f}(T;\phi_{0})\equiv \VEV{{\cal
O}_{f}(T;\phi_{0})}~,~T>0\eqn\sabbioneta$$
can be evaluated with the same method as above. The result is
(up to an irrelevant factor)
$$\Gamma_{f}(T;\phi_{0})=S^{[a_{1}}_{~a_{1}}(T;\phi_{0})S^{a_{2}}_{a_{2}}
(T;\phi_{0})\cdots S^{a_{f}]}_{a_{f}}(T;\phi_{0})\eqn\casalmaggiore$$
where the square brackets denote the complete antisymmetrization;
for f=2, for example, ~\casalmaggiore~ reads as
 ~$\Gamma_{2}=(TrS)^{2}-Tr(S^{2})$. The interpretation
of the observables ~${\cal O}_{f}$~is as follows. At time t=0 the
operator~${\bar c}_{a_{f}}(0)\cdots {\bar c}_{a_{1}}(0)$~ creates a
state with  f ghosts (or better with  f Jacobi-fields) from the
 vacuum.
In the geometric interpretation of the theory this state corresponds
to a  f-form ~ $d\phi^{a_{1}}\wedge d\phi^{a_{2}}\wedge\cdots
d\phi^{a_{f}}$,\ie, to a  f-dimensional volume. Hence ~$\Gamma_{f}$
contains information about the rate of growth of  f-dimensional
volume elements in tangent space.
\par
The expectation values ~$\Gamma_{f}$~ are related to the Lyapunov
exponents ~$\lambda_{i}^{(1)}(\phi_{0})$ as follows. Consider first
$$\Gamma_{1}(t;\phi_{0})=Tr S(t;\phi_{0})\eqn\piadena$$
and assume that the initial point ~$\phi_{0}$~ gives rise to a
periodic trajectory with period
~$\tau$,\ie, $\phi^{a}(t)=\phi^{a}(t+\tau)$. Consequently the matrix ~
$M$~ of eq.\baleari~ is periodic too, and Floquet theory
\Ref\Cro{See, for example, J.Cronin, {\it "Differential equations"},\nextline
Marcel Dekker, New York, 1980} tells us that ~$S$~ can be
written as
$$S(t)=P(t)exp (Rt)\eqn\brescello$$
where ~$P$~ is a periodic matrix, ~$P(t)=P(t+\tau)$, and
~$R$~ is a constant one. Because \break $S(0)=P(0)=1$ we have
~$S(\tau)=exp(R\tau)$. Let us then diagonalize ~$R$;
its eigenvalues~$\rho_{a},~a=1\cdots 2n$, are the characteristic
exponents of ~$M(t)$ and ~$exp(\rho_{a}t)$~ the corresponding
Floquet multipliers.
In the eigenvector basis of ~$R$~ we have (suppressing the
argument ~$\phi_{0}$)
$$S^{a}_{b}(t)=P^{a}_{b}(t)exp (\rho_{a}t)\eqn\boretto$$
We assume the ordering
$${\it Re}\rho_{1}\geq{\it Re}\rho_{2}\geq\cdots\geq{\it Re}
{\rho_{2n}}\eqn\guastalla$$
If ~${\it Re}\rho_{1}$~ is strictly larger than ~${\it Re}
\rho_{2}$, the large-t behaviour of ~$\Gamma_{1}$ is
$$\Gamma_{1}(t;\phi_{0})\sim P^{1}_{~1}exp(\rho_{1}t)\eqn\poviglio$$
Hence
$$g_{1}(\phi_{0})\equiv \limsup_{t\rightarrow\infty}{1\over
t}ln\Gamma_{1}(t;\phi_{0})=\rho_{1}\eqn\reggio$$
exists and coincides with the Lyapunov exponent
~$\lambda_{1}^{(1)}(\phi_{0})$.

Because ~$S(\tau)$~ is a real symplectic matrix, its eigenvalues
appear as 4-tuples: if ~$\mu$~ is a complex eigenvalue
then \refmark{14}
 ~$\mu^{\star}$,~${1\over\mu}$ and ~${1\over\mu^{\star}}$
are eigenvalues too (not necessarily different from ~$\mu$).
Since ~$S(\tau)=exp(R\tau)$ this means that if ~$\rho$~ is a
characteristic exponent, then also ~$\rho^{\star}$,$-\rho$~ and
$-\rho^{\star}$~ are characteristic exponents. Let us write
~$\rho_{a}\equiv l_{a}+i\omega_{a}$~ with
~$l_{a}$~ and ~$\omega_{a}$~ real. As for the relative magnitude
of the numbers ~${\it Re}\rho_{a}=l_{a}$~ two cases have to be
distinguished. If some ~$\rho_{a}$~ has a nonvanishing imaginary
part,~$\omega_{a}\not=0$, then ~$\rho_{a}^{\star}$~ is different
from ~$\rho_{a}$~ and consequently there exist two exponents with the
same real part (they give rise to equal contributions to the Lyapunov
exponents). On the other hand, if \break ~$\omega_{a}=0$~, the exponent
~$\rho_{a}$~ is real and generically there will be no other
exponent with the same real part. In this case ~$\Gamma_{1}$, say, is
dominated by a single real multiplier,\ie,~$\rho_{1}$. If however,
~$\rho_{1}\equiv l_{1}+i\omega_{1}$~ and
~$\rho_{2}\equiv l_{1}-i\omega_{1}$~
are a complex conjugate pair, eq.\poviglio~ is replaced by
$$\Gamma_{1}(t;\phi_{0})\sim
\bigr[R^{1}_{1}(t)~e^{i\omega_{1}t}+R^{2}_{2}e^{-i\omega_{1}t}\bigl]
e^{l_{1}t}\eqn\cogozzo$$
\par
From ~\casalmaggiore~ with ~\boretto~ we obtain for the higher
correlation functions
$$\Gamma_{f}(t;\phi_{0})=P^{[a_{1}}_{~a_{1}}(t)\cdots
P^{a_{f}]}_{a_{f}}(t)~exp\bigl[(\rho_{a_{1}}+\rho_{a_{2}}+\cdots
+\rho_{a_{f}})t\bigr]\eqn\cicognara$$
Due to the antisymmetrization, the indices ~$a_{j}$ of the ~$\rho$'s
in the exponential must all be different. Because of the ordering
~\guastalla~ this implies that for ~$t\rightarrow\infty$
$$\Gamma_{f}(t;\phi_{0})\sim p(t)~exp\bigl[(\rho_{1}+\rho_{2}+\cdots
+\rho_{f})t\bigr]\eqn\smatteo$$
for some ~$\tau $-periodic function p(t). As it stands ~\smatteo~ is
correct only if the real part of the last eigenvalue,
~$l_{f}={\it Re}\rho_{f}$, is strictly larger than the real part of the
 following
eigenvalue,~$l_{f+1}$. If, for some reason \foot{e.g., because
~$\rho_{f}$~ and ~$\rho_{f+1}$ ~form a complex conjugate pair.}
$l_{f}=l_{f+1}$, the asymptotic formula consists of two terms,
$$\eqalign{\Gamma_{f}(t;\phi_{0})\sim p(t)~exp\bigl[(\rho_{1}+\rho_{2}+\cdots
\rho_{f-1}\ + & \ \rho_{f})t\bigl]+\cr
\ {\widetilde p}(t) & \ ~exp\bigl[(\rho_{1}+
\rho_{2}+\cdots+\rho_{f-1}+\rho_{f+1})t\bigr]\cr}\eqn\roncadello$$
or even more terms if ~$\rho_{f}$~ is degenerate with ~$\rho_{f+2}$,
$\rho_{f+3}$ , $\cdots$ ,etc.
In any case
$$g_{f}(\phi_{0})\equiv \limsup_{t\rightarrow \infty}
{1\over t}ln\Gamma_{f}(t;\phi_{0})=\sum_{i=1}^{f}l_{i}
(\phi_{0})\eqn\pomponesco$$
is the sum of the f largest real parts of the eigenvalues
~$\rho_{a}$.
\par
In the introduction we mentioned already that the leading Lyapunov
exponent governing the evolution of  f-forms,
~$\lambda_{1}^{(f)}(\phi_{0})$, is related to the higher Lyapunov
exponents for one forms, ~$\lambda^{(1)}_{i}(\phi_{0})$, according to
~$$\lambda^{(f)}_{1}(\phi_{0})=\sum_{i=1}^{f}\lambda_{i}^{(1)}
(\phi_{0})\eqn\dosolo$$
This is exactly the relation we found in eq.~\pomponesco. The
correlation function~$\Gamma_{f}$ is the expectation value of the
operator ~${\cal O}_{f}$~ which creates and destroys a 
f-dimensional "parallelotope", and hence ~$g_{f}$~ describes the rate of
exponential growth of f-dimensional volume elements in tangent
space: ~$\lambda^{(f)}_{1}(\phi_{0})=g_{f}(\phi_{0})$.
Once ~$g_{f}$ is known for all values of f, the system of equations
~\pomponesco~ can be solved for
~$l_{i}(\phi_{0})=\lambda_{i}^{(1)}(\phi_{0})$~ in order to obtain the
higher Lyapunov exponents for one forms,\ie,
~$\lambda_{i}^{(1)}(\phi_{0})$.
\par
So far we have shown that the correlation functions
~$\VEV{{\cal O}(t;\phi_{0})}$~ encapsulate the information
about all the Lyapunov exponents related to a fixed trajectory,
namely the one starting at ~$\phi_{0}$~ a time t=0. If we restrict
~$\phi_{0}$~ to a region in phase-space of connected stochasticity
(excluding  regions of regular motion)
the Lyapunov exponents are independent
of ~$\phi_{0}$~ and we may equally well extract them from the
observable~\viadana~ with the delta-function fixing the initial
point omitted, but under a path-integral which is
over a restricted class of trajectories only. (As was shown by
Oseledec\refmark{13} it is not really necessary to insist on closed
trajectories.)
As we shall see in the next section, expectation
values of this type are closely related to the partititon function
of the superhamiltonian ~${\widetilde{\cal H}}$.

\chapter{PARTITION FUNCTIONS OF THE SUPERHAMILTONIAN}
Obviously the superhamiltonian ~${\widetilde{\cal H}}$
of eqs.~\citroen,~\renault~ does not mix forms of different
degree (ghost number). Therefore it makes sense to consider
~${\widetilde{\cal H}}_{p}$, the restriction of
${\widetilde{\cal H}}$~to the space of homogeneous
p-forms\foot{For an analogous construction in supersymmetric
quantum mechanics see ref[18]\REF\hal{M.Claudson and M.B.Halpern,
Ann.Phys. 166 (1985) 33;\nextline
R.Graham, Phys.Lett.109A (1985) 436}.},
which is spanned by the generalized densities
of the type ~\mann. Let us evaluate the partition
function
$$Z_{p}(T)=Tr\bigl[exp(-i{\widetilde {\cal H}}t)\bigr]=\sum_{\alpha}
<\chi^{\alpha}_{p}\vert exp(-i{\widetilde {\cal H}}_{p}t)\vert \chi^{\alpha}
_{p}>\eqn\parma$$
where ~$\bigl\{\ket{\chi^{\alpha}_{p}}\bigr\}$~is a basis of
p-forms and ~$\bigl\{\bra{\chi^{\alpha}_{p}}\bigr\}$~ the dual basis
of  p-vectors (antisymmetric contravariant tensors of degree p).
In component notation the completness relation reads
$$\sum_{\alpha}\chi^{\alpha}_{p}(\phi)_{a_{1}\cdots a_{p}}
\chi^{\alpha}_{p}(\phi^{\prime})^{\star b_{1}\cdots b_{p}}=
{\bf
\delta}^{(2n)}(\phi-\phi^{\prime}){\bf\delta}^{[b_{1}}_{a_{1}}
\cdots{\bf\delta}^{b_{p}]}_{a_{p}}\eqn\bologna$$
From ~\audi ~and ~\bastia~ we obtain for the matrix element
of ~$exp(-i{\widetilde H}_{p}t)$
$$\eqalign{\bra{\phi,c}exp(-i{\widetilde H}_{p}\ t & \ )\ket{\chi^{\alpha}
_{p}}  =
<\Phi^{-1}_{cl}(t;\phi),S^{-1}(t;\phi)c\vert\chi^{\alpha}_{p}>=\cr
\ = & \ {1\over p!}\chi^{\alpha}_{p}\bigl(\Phi^{-1}_{cl}(t;\phi)\bigr)
_{a_{1}\cdots a_{p}}S^{-1}(t;\phi)^{a_{1}}_{b_{1}}\cdots
S^{-1}(t;\phi)^{a_{p}}_{b_{p}}c^{b_{1}}\cdots c^{b_{p}}\cr}\eqn\pisa$$
The partition function~\parma~is obtained by stripping off the ghosts
from~\pisa~, contracting with the dual basis and integrating over
~$\phi$~. Exploiting ~\bologna~ one finds (up to an unimportant
constant)
$$\eqalign{Z_{p}(T)\ = & \ \int~d\phi~{\bf \delta}(\phi-\Phi^{-1}_{cl}
(T;\phi))S^{-1}(T;\phi)^{[a_{1}}_{a_{1}}\cdots
S^{-1}(T;\phi)^{a_{p}]}_{a_{p}}\cr
\ \sim & \
\int~d\phi~{\bf\delta}(\Phi_{cl}(T;\phi)-\phi)~S(T;\phi)^{[a_{1}}_{a_{1}}
\cdots S(T;\phi)^{a_{2n-p}]}_{a_{2n-p}}\cr}\eqn\livorno$$
In the second line we used the identity~[A.55] (which is derived in
the appendix) and the fact
that the Jacobi matrix is unimodular. As it was to be expected,
~$Z_{p}(T)$~ receives contributions only from closed trajectories of
period T.
\par
Next we show that the partition functions in the p-form and the
(2n-p)-form sector coincide:
$$Z_{p}(T)=Z_{2n-p}(T)\eqn\firenze$$
The reason is that there exists a duality operation~$\star$~
which maps p-forms on (2n-p)-forms and which commutes with
~${\widetilde{\cal H}}$. On an arbitrary p-form
~$\chi$~ it acts as
$$({\star}\chi)_{a_{p+1}\cdots a_{2n}}\sim \epsilon_{a_{1}\cdots
a_{2n}}\omega^{a_{1}b_{1}}\cdots \omega^{a_{p}b_{p}}\chi_{b_{1}\cdots
b_{p}}\eqn\siena$$
This kind of ~$\star$~ operator is analogous to the Hodge operator
on Riemannian manifolds. The ~$\star$-operation commutes with
~${\widetilde{\cal H}}\equiv -il_{h}$~ because the Lie-derivative
along the hamiltonian vector field of both the ~$\varepsilon$-tensor
and of the ~$\omega^{ab}$~ vanishes.
(The equation ~$l_{h}\varepsilon_{a_{1}\cdots a_{p}}=0$ is the component
form of eq.~\infi.) Hence the spectra of ~${\widetilde{\cal H}}_{p}$~
and ~${\widetilde{\cal H}}_{2n-p}$~ coincide, which is analogous to
the well-known Poincar\'e duality for the Laplacian.
From~\livorno~ with ~\firenze~ and ~\casalmaggiore~ it follows that
$$\eqalign{Z_{f}(T) \ = & \ \int
d\phi~{\bf\delta}(\Phi_{cl}(T;\phi)-\phi)~S(T;\phi)^{[a_{1}}_{~a_{1}}\cdots
S(T;\phi)^{a_{f}]}_{a_{f}}\cr
\ = & \ \int d\phi
~{\bf\delta}(\Phi_{cl}(T;\phi)-\phi)~\Gamma_{f}(T;\phi)\cr}\eqn\arezzo$$
Thus we find that the ratio ~${Z_{f}(T)\over Z_{0}(T)}$~ can be
interpreted as the average of ~$\Gamma_{f}(T)$~ over all closed
trajectories of (not necessarily primitive) period T.
Because ~$\Gamma_{f}$=$\VEV{{\cal O}_{f}}$~ \break with ~${\cal O}_{f}$ given
in ~\viadana~, we easily could write down a path-integral
representation for ~$Z_{f}$~ by combining~\arezzo~ with~\milano.
Note also that
$$Z_{0}(T)=\int~d\phi~{\bf\delta}(\Phi_{cl}(T;\phi)-\phi)\eqn\pistoia$$
"counts" the number of initial points which lead to a closed
trajectory of period T. Of course these points are  infinite in
number. What we should do in ~\pistoia~, and in all the other ~$Z_{f}(T)$,
is to factor out the equivalence relation which relates two
points which are on the same trajectory. This work is in progress
\Ref\thac{E.Gozzi, M.Reuter, W.D.Thacker, work in progress}. After
having factored
out the equivalence relation, the new ~$Z_{0}(T)$~ would count
the number of closed orbits of period T and that implies
that from this  new ~$Z_{0}(T)$
we should be able to get the {\it topological entropy} of the system
\Ref\proc{For a short review see: I. Procaccia, Nucl.Phys.B (Proc.
\break Suppl.)
2 (1987) 527}. Anyhow the reader should not
be worried about the "infinity" produced by ~$Z_{p}(T)$~
because the physical quantitities which we will consider
are always given by ratios of ~$Z_{p}(T)$~ and in these ratios
the infinity  cancels out.
\par
Until now we were purposely vague about the domain of the ~$\phi$~
integration in eq.~\arezzo. If the trace in ~\parma~ is over p-forms
defined on the full phase-space ~${\cal M}_{2n}$~then clearly the
integration in ~\arezzo~ is over all of ~${\cal M}_{2n}$. However,
in general one would like to discuss the chaoticity properties
of a system in different regions of phase-space separately,
and, most importantly one would like to work at fixed energy
E. The (2n-1)-dimensional energy hypersurface ~${\cal M}_{2n-1}
(E)$~ is the subspace of ~${\cal M}_{2n}$ on which ~$H(\phi)=E$. If the
initial condition of the classical path-integral are fixed such that
the initial point ~$\phi(0)$~ lies on ~${\cal M}_{2n-1}(E)$, then
the dynamics is such that ~$\phi(t>0)$~ is still on~${\cal M}_{2n}$.
Correspondingly, if a zero-form ~$\varrho(\phi,t=0)$~ has support
on ~${\cal M}_{2n-1}(E)$~only, this property is conserved under the
time evolution. This is not sufficient, however. We also have to make
sure that p-forms on ~${\cal M}_{2n-1}(E)$~ evolve into p-forms on
~${\cal M}_{2n-1}(E)$. A p-form on ~${\cal M}_{2n-1}(E)$~ is a tensor
which has no components in the direction perpendicular to the energy
hypersurface. Loosely speaking,
the exterior algebra on ~${\cal M}_{2n-1}(E)$~is obtained by putting
to zero the component of ~$d\phi^{a}$~ normal to the energy
hypersurface:~$\partial_{a}H(\phi)d\phi^{a}=0$. Therefore, defining
~$N(t)\equiv \partial_{a}H(\phi(t))c^{a}(t)$, we have to constrain
the path-integration to the subspace with ~$N(t)=0$. In the second
of ref.[10] we have shown that the charge ~$N$~ is conserved under
the time evolution: $\bigl[N,{\widetilde{\cal H}}\bigr]=0$. In fact,
~$N$~ is the difference between the supersymmetry generator and the
BRS-generator: $N=Q_{H}-Q$. This implies that, imposing ~$N(t=0)=0$,
guarantees that ~$N(t)$=0 at any later time. As a consequence, if
~${\widetilde{\varrho}}(\phi,c,t=0)$~ is a tensor on ~${\cal M}_{2n-1}
(E)$, (\ie, if it does not contain any  factor of ~$N$), also
~${\widetilde\varrho}(\phi,c,t)$~at ~$t>0$~ is a tensor on ~
${\cal M}_{2n-1}(E)$. Thus we can consistently truncate the classical
path-integral to the energy hypersurface. In particular we may define
partition functions ~$Z_{p}(T;E)$~ as in eq.\parma~ but with
~$\chi^{\alpha}_{p}$~ a complete set of p-forms on ~${\cal
M}_{2n-1}(E)$. The next step is to develop a full symplectic and
coordinate free formalism. This implies that we
will have to decrease the dimension of ~${\cal M}_{2n-1}(E)$~
by one unit to go to an  even-dimensional subspace of ~${\cal M}_{2n}$  .
This is done via the so-called
Batalin-Fradkin-Vilkovisky method\Ref\bat{For a review and references see
M.Henneaux,
Phys.Rep.126 (1985) 1}(BFV-formalism)
which implies the introduction of further auxiliary fields and further
ghosts. The details of this will be presented elsewhere
\refmark{19}.
\par
Let us now go back to~$Z_{f}(T)$. From the asymptotic 
~${T\rightarrow\infty}$~ behaviour
of ~$Z_{f}(T)$~ we define the generalized Lyapunov
exponents ~$\Lambda_{f}$, $1\le f\le 2n$~, according to
$${Z_{f}(T)\over Z_{0}(T)}\sim exp \bigl[(\Lambda_{1}+\Lambda_{2}+
\cdots+\Lambda_{f})T\bigr]\eqn\urbino$$
This is the deterministic analogue of eq.(A.57) for stochastic systems
which we will find in  appendix A. Formally the exponent
~$\Lambda_{1}$, say, is defined as in eq.\rosita~with the only
difference that the ensemble average ~$\VEV{\cdot}$~ is not taken with
respect to the stochastic measure but with the deterministic one.
Because we formulated also the deterministic systems in a
path-integral language, this correspondence becomes particularly
transparent. Note that generically (for a deterministic system) there
is no simple relation between the generalized exponents
~$\Lambda_{i}$~ and the ordinary ones, $\lambda_{i}^{(1)}$.
In order to obtain the former, one averages the monodromy
matrix ~$S(T)$~ for many paths of length T and sends T to infinity
afterwards, whereas the latter one is obtained from the large-T
behaviour of S(T) on a single trajectory.
\par
We now briefly comment on the eigenvalue
problem of the superhamiltonian ~${\widetilde{\cal H}}$~ and the
time evolution operator ~$exp(-i{\widetilde{\cal H}}t)$. It is at this
point that we encounter the most important differences between
stochastic systems and classical hamiltonian systems. In the former
case (see appendix A) the superhamiltonian is a second order
Schroedinger operator, in the latter it is the first
order Lie-derivative operator~${\widetilde{\cal H}}=-il_{h}$.
Let ~$\chi(\phi,c)$~ be an eigenfunction of ~${\widetilde{\cal H}}$~
in the p-form sector so that
$$exp(-i{\widetilde{\cal H}}t)\chi(\phi,c)=exp(-i{\widetilde{\cal
E}}t)\chi(\phi,c)\eqn\perugia$$
for some constant~${\widetilde{\cal E}}$. Using eq.~\pisa~
we immediately see that the components
~$\chi_{a_{1}\cdots a_{p}}(\phi)$~ satisfy the eq.
$$\chi_{a_{1}\cdots a_{p}}\bigl(\Phi_{cl}(t;\phi)\bigr)
S(t;\phi)^{a_{1}}_{b_{1}}\cdots S(t;\phi)^{a_{p}}_{b_{p}}=
exp(+i{\widetilde{\cal E}}t)\chi_{b_{1}\cdots
b_{p}}(\phi)\eqn\foligno$$
On the LHS of this equation we recognize the usual tensorial
transformation law under the Hamiltonian flow. It affects
the eigenfunctions ~$\chi$~ only via the overall factor
~$exp(i{\widetilde{\cal E}}t)$. Let us look at p=1 in more detail,
where the relation is
$$\chi_{a}\bigl(\Phi_{cl}(t;\phi)\bigr)S(t;\phi)^{a}_{b}=exp
(i{\widetilde{\cal E}})\chi_{b}(\phi)\eqn\orte$$
and it has to hold for all ~$\phi$~ and all t if ~$\chi$~ is an
eigenfunction.
Let us assume we pick a point ~$\phi_{0}$~ on ~${\cal M}_{2n}$~
which is the initial point of a closed trajectory of duration ~$\tau$:
$\Phi_{cl}(\tau;\phi_{0})=\phi_{0}$. Let ~$S(\tau;\phi_{0})$~ be the
the Jacobi matrix evaluated at $t=\tau$\foot{This is what is called
in the literature\Ref\Giv{V.Arnold,  A.B.Givental,  "{\it Dynamical
systems IV"},\nextline
Springer-Verlag, New York, 1990} "monodromy matrix" of this loop.}.
If we evaluate ~\orte~ for the special values ~$\phi=\phi_{0}$~ and
~$t=\tau$~, we find
$$\bigl[S(\tau;\phi_{0})^{a}_{b}-exp(i{\widetilde{\cal
E}}\tau)\delta^{a}_{b}\bigr]\chi_{a}(\phi_{0})=0\eqn\roma$$
\ie, that ~$exp(i{\widetilde{\cal E}}\tau)$~ is an eigenvalue of the
monodromy matrix provided that ~$\chi_{a}(\phi_{0})\not=0$.
This shows that the possible values of ~${\widetilde{\cal E}}$~ are
closely related to the periods and the Floquet multipliers
of closed orbits.
\par
As a final remark we mention that in a previous paper
(the fourth of ref.[10]) we had shown
that the alternating sum
$$Z(T)\equiv STr\bigl[exp(-i{\widetilde {\cal H}}t)\bigr]=\sum_{p=0}^{2n}
(-1)^{p}Z_{p}(T)\eqn\agrigento$$
is a topological invariant (the Euler number of ~${\cal M}_{2n}$)
and has no dynamical significance therefore. It was proven in fact that
the alternating sum~\agrigento~ is invariant
under deformation of the hamiltonian vector fields. On the other hand, we
saw that the individual ~$Z_{p}$'s  are not invariant under such
deformations  and therefore can be used to characterize certain
properties of the dynamics.
\chapter{CONCLUSIONS}
To summarize we can say that, looking back at eqs.~\viadana,~\sabbioneta,
~\pomponesco for the ordinary Lyapunov exponents and at
~\urbino~and ~\parma for the generalized ones,
both exponents admit a very simple and natural representation in terms of
classical path-integrals.The representation of the
Lyapunov exponents as expectation values of some observables allows, for
example, then for a perturbative calculations of them in the same manner as
it is usually done in field-theory. Viceversa the representation
via eq.~\urbino~ allows the use of spectral methods to calculate then.
The supersymmetry of the path-integral is crucial in this
context: it relates the "bosonic" dynamics of the trajectories
~$\phi(t)$~ to the evolution of its "fermionic" superpartner,
the Jacobi-field ~$c(t)$ and because of this relation it was natural to expect
that information on the dynamics of the Jacobi-fields
could be extracted from some objects containing only the dynamics of
the standard-phase-space variables as ~$Z_{f}$~is. This supersymmetry
will also produce\refmark{16} semplifications in the perturbative calculations
in the same way as it does in standard  field theory.
\par
A lot of work remains to be done in order to extract the dependence
on the energy of the ordinary (and generalized)
Lyapunov exponents and  of the various entropy-like quantities
that are associated to them. This work is in progress
\refmark{19} together with an understanding of the relation
of our formalism with the thermodynamic formalism of Ruelle
\Ref\Rue{
D.Ruelle, Inventiones math.34 (1976) 231;\nextline
D.Ruelle, {\it Thermodynamic Formalism},
Addison-Wesley, Reading, \break Mass., 1978;\nextline
D.Ruelle, Com. Math.Phys.125 (1989) 239;\nextline
Y.Oono and Y.Takahashi, Prog. Theor.Phys.63 (1980) 1804;\nextline
Y,Takahashi and Y.Oono, Prog.Theor.Phys. 71 (1984) 851;\nextline
B.Eckhardt and P.Cvitanovic, Jour. Phys.A,  24 (1991) L237} on
which we briefly comment in appendix B.
\Appendix{A}
In this appendix we use the tools of supersymmetric quantum
mechanics in order to discuss the generalized Lyapunov exponents for
{\it stochastic systems}. The reader should compare the various steps
of the derivation with their classical counterparts described in the
main body of the paper.
\par
Let us  consider a stochastic process on an N-dimensional, metrically
flat configuration space with (local) coordinates ~$x_{i}$~$i=1,\cdots,
N$. The dynamics of the random variables ~$x_{i}(t)$~ is given by the Langevin
equation
$${\dot x}_{i}(t)=
-\partial_{i}U\bigl(x(t)\bigr)+\eta_{i}(t)\eqn\bella$$
where ~$U(x)$~ is a smooth potential and ~$\eta_{i}(t)$~ is a white
noise :
$$\VEV{\eta_{i}(t)}=0,~~\VEV{\eta_{i}(t)\eta_{j}(t^{\prime})}=\delta
_{ij}\delta(t-t^{\prime})\eqn\brutta$$
It is well known\refmark{7}
that the stochastic correlations\break~$\VEV{x_{\eta}\cdots x_{\eta}}$~
derived from~\bella~ can be obtained from a supersymmetric
generating functional of the form:
$$Z_{susy}=\int {\cal D}x{\cal D}\psi{\cal D}{\bar\psi}~
exp\bigl(-\int dt {\it L}_{susy}+source~terms\bigr)\eqn\discreta$$
where
$${\it L}_{susy}={1\over 2}{\dot x}_{i}^{2}+{1\over 2}\bigl(\partial
_{i}U\bigr)^{2}+{\bar\psi}_{i}\bigl[\partial_{t}\delta_{ij}+\partial_{i}
\partial_{j}U(x)\bigr]\psi_{j}\eqn\mediocre$$
The ~$x_{i}$ are commuting variables while ~$\psi_{i}$~ and
~${\bar\psi}_{i}$ are Grassmannian ones. The supersymmetry
transformations under which
\mediocre~ is invariant are given by:
$$\eqalign{ \delta x_{i} \ = & \ -\epsilon\psi_{i}+{\bar\epsilon}
{\bar\psi}_{i}\cr
\delta\psi_{i} \ = & \ {\bar\epsilon}(-{\dot x}_{i}+\partial_{i} U)\cr
\delta{\bar\psi}_{i} \ = & \ {\epsilon}({\dot
x}_{i}+\partial_{i}U)\cr}\eqn\suss$$
In a Schroedinger picture formulation of the supersymmetric quantum
\break mechanics\refmark{16,18} defined by ~\mediocre~ the states ~$\ket{\Phi}$
are
described by wave functions
$$\Phi(x_{i},\psi_{i})\equiv <x_{i},\psi_{i}\vert\Phi>\eqn\jaki$$
depending on the variables ~$x_{i},\psi_{i}$. The operators
~${\hat x}_{i}$~ and ~${\hat\psi}_{i}$~ act on ~${\Phi}(x,\psi)$~ by
multiplication and their conjugate momenta by differentiation:
$${\hat p}_{i}={\partial\over\partial x_{i}}\equiv\partial_{i}~~,
~~{\hat{\bar\psi}}_{i}={\partial\over\partial\psi_{i}}\eqn\solit$$
Eq.\mediocre~ gives rise to the (Weyl ordered) superhamiltonian:
$$H_{susy}=H_{B}+H_{F}\eqn\carmen$$
where the "bosonic" part is:
$$H_{B}\equiv -{1\over 2}\partial^{2}+{1\over
2}\partial_{i}U\partial_{i}U,~~~~\partial^{2}\equiv
\partial_{i}\partial_{i}\eqn\silvio$$
and the "fermionic" part is:
$$\eqalign{H_{F} \ \equiv & \ {1\over 2}\partial_{i}\partial_{j}U
\bigl({\hat{\bar\psi}}_{i}{\hat\psi}_{j}-{\hat\psi}_{j}{\hat{\bar\psi}}_{i}
\bigr)\cr
\ = & \ {1\over 2}
\partial^{2}U(x)-\partial_{i}\partial_{j}U(x)\psi_{j}{\partial\over\partial
\psi_{i}}\cr}\eqn\giuliana$$
In terms of the supercharges
$$\eqalign{ Q \ = & \
\bigl[-\partial_{i}+\partial_{i}U(x)\bigr]\psi_{i}\cr
{\bar Q} \ = & \ \bigl[\partial_{i}+\partial_{i}U(x)\bigr]{\partial
\over\partial\psi_{i}}\cr}\eqn\tina$$
we have
$$H_{susy}={1\over 2}\bigl[ Q,{\bar Q}\bigr]\eqn\paola$$
\par
In this operatorial formalism\refmark{16} a generic wave function
~$\Phi(x,\psi)$~ possesses an expansion of the form
$$\Phi(x,\psi)=\sum_{q=0}^{N}{1\over
q!}\Phi^{(q)}_{k_{1}\cdots k_{q}}(x)\psi^{k_{1}}\cdots
\psi^{k_{q}}\eqn\rico$$
which is reminiscent of the expansion of an inhomogeneous
differential form in a basis \break $dx^{k_{1}}\land \cdots \land
dx^{k_{q}}$. We say that ~$\Phi$~ has ghost number "p" if on the RHS
of~\rico~ only the term with q=p is different from zero
(homogeneous form of degree p):
$$\Phi(x,\psi)={1\over p!}\Phi_{k_{1}\cdots k_{p}}(x)\psi^{k_{1}}\cdots
\psi^{k_{p}}\eqn\belletti$$
The vacuum of the fermionic Fock space, defined by ~${\hat\psi}_{i}\ket
{vac}=0$,~ is represented by a wave function of ghost number n:
$$<x,\psi\vert vac>=\Phi_{vac}(x){\bf\delta}(\psi)\eqn\vuoto$$
where the Grassmannian delta-function
$${\bf\delta}(\psi)=\psi^{1}\psi^{2}\cdots \psi^{N}\eqn\chissa$$
can be visualized as describing a completely filled "Dirac sea".
Multiparticle states are obtained from ~$\ket{vac}$~by acting on
~${\bf\delta}(\psi)$~ with the "creation operator"
~${\hat{\bar\psi}}_{i}={\partial\over\partial\psi_{i}}$. A state
containing f "particles" has a wave function with ghost number
N-f:
$$<x,\psi\vert \Phi_{f}>=\Phi_{k_{1}\cdots k_{f}}(x)
{\partial\over\partial\psi_{k_{f}}}\cdots {\partial\over\partial
\psi_{k_{1}}}{\bf\delta}(\psi)\eqn\forse$$
The time evolution of the states ~$\Phi(x,\psi,t)$~is governed
by the Schroedinger equation
$$H(x,{1\over i}{\partial\over\partial x},\psi, {\partial\over\partial \psi})
\Phi(x,\psi,t)=-\partial_{t}\Phi(x,\psi,t)\eqn\schroede$$
It has the formal solution ($t>0$)
$$\Phi(x,\psi,t)=\int d^{N}x_{0}~d^{N}\psi_{0}~
{\bf K}(x,\psi,t\vert
x_{0},\psi_{0},t_{0})\Phi(x_{0},\psi_{0},t_{0})\eqn\mario$$
The evolution kernel ~${\bf K}$~ is a solution of the Schroedinger equation
\schroede~ with initial condition
$${\bf K}(x,\psi, t_{0}\vert x_{0},\psi_{0},t_{0})={\bf\delta}^{N}(x-x_{0})
{\bf\delta}^{N}(\psi-\psi_{0})\eqn\maria$$
Its path integral representation involves the lagrangian~\mediocre:
$${\bf K}(x_{2} ,\psi_{2},t_{2}\vert
x_{1},\psi_{1},t_{1})=  \int {\cal D}x(t)~{\cal D}\psi(t)~{\cal D}{\bar\psi}
(t)~exp\bigl\{-\int^{t_{2}}_{t_{1}}dt {\it L}_{susy}\bigr\}\eqn\ennio$$
The boundary conditions are
~$x(t_{1,2})=x_{1,2},\psi(t_{1,2})=\psi_{1,2}$~ and
~${\bar\psi}(t_{1,2})$~ is integrated over. (For details see
ref.[16,12])
Writing the above kernel as
$$\eqalign{{\bf K}(x_{2},\psi_{2},t_{2}\vert x_{1},\psi_{1},t_{1})
\ = & \
\int {\cal D}x(t)~exp\bigl\{-\int_{t_{1}}^{t_{2}}dt\bigl[{1\over 2}
{\dot {x}}_{i}^{2}+{1\over 2}(\partial_{i}U)^{2}\bigr]\bigr\}\cdot\cr
\ {\cdot} & \ {\bf K}_{F}(\psi_{2},t_{2}\vert \psi_{1},t_{1};[x])\cr}
\eqn\marto$$
with
$$ {\bf K}_{F}(\psi_{2},t_{2}\vert \psi_{1},t_{1};[x])\equiv
\int {\cal D}\psi~{\cal D}{\bar\psi}~exp\bigl\{-\int^{t_{2}}_{t_{1}}
dt
{\bar\psi}_{i}\bigl[\partial_{t}\delta_{ij}+\partial_{i}\partial_{j}U(x)\bigr]
\psi_{j}\bigr\}\eqn\marton$$
we can explicitly evaluate the fermionic kernel ~$K_{f}$~which is a
functional of the bosonic path ~$x_{i}(t)$. Following the treatment
of ref.[12], we first perform the (unconstrained) ~${\bar\psi}$
-integration in eq.\marton~ which leads to
$${\bf K}_{F}(\psi_{2},t_{2}\vert \psi_{1},t_{1};[x])=\int {\cal D}
\psi~{\bf\delta}[\bigl(\partial_{t}\delta_{ij}+\partial_{i}
\partial_{j}U(x)\bigr)
\psi_{j}]\eqn\bando$$
Obviously only the solutions of the equation
$${\dot\psi}_{i}=-\partial_{i}\partial_{j}U(x(t))\psi_{j}\eqn\jacob$$
obeying the boundary conditions ~$\psi(t_{1,2})=\psi_{1,2}$~
contribute to ~${\bf K}_{F}$~. This is a rather remarkable fact, because
eq.\jacob~ is precisely the Jacobi eq. pertaining to the Langevin
equation~\bella~ (\ie, if ~$x_{i}(t)$ is a solution
of ~\bella, then the variation~$\delta x_{i}(t)\equiv {\psi}_{i}
(t)$ is a solution of~\jacob). The explicit solution of eq.\jacob~
reads
$$\psi_{i}(t)=S_{ij}(t;[x])\psi_{j}(t_{1})\eqn\merlo$$
with the Jacobi matrix
$$S(t;[x])={\widehat T} exp \bigl\{-\int^{t}_{t_{1}}dt^{\prime}
M\bigl(x(t^{\prime})\bigr)\bigr\}\eqn\passero$$
(${\widehat T}$~ denotes the time ordering operator) and where
$$M_{ij}\bigl(x(t)\bigr)\equiv \partial_{i}\partial_{j}
U\bigl(x(t)\bigr)\eqn\uva$$
$S$ is a solution of the matrix equation ~${\dot S}=-MS$~
with ~$S(t_{1})=1$. We obtain for the kernel ~\bando~
$$\eqalign{ {\bf K}_{F}(\psi_{2} \ , & \ t_{2}\vert \psi_{1},t_{1};[x])
=\cr
\ = & \ \int {\cal
D}\psi~{\bf\delta}\bigl[\psi(t)-S(t;[x])\psi_{t_{1}}\bigr]~det\bigl[
\partial_{t}\delta_{ij}+\partial_{i}\partial_{j}U(x)\bigr]\cr}
\eqn\grano$$
In order to give a well-defined meaning to the determinant, we have
to specify a discretization scheme for the functional integral.
Because the Hamiltonian in ~\carmen~ was chosen to be Weyl
ordered, we must use the mid point rule
for the discretization.\Ref\sak{B.Sakita, {\it "Quantum theory of many
variable systems and fields"},\nextline
World Scientific Publ., Singapore, 1985}
For this discretization, it is known that
\Ref\goz{This has been shown by many authors and it is
summarized in \break E.Gozzi, Phys.Rev.D28 (1983) 1922 }
$$det\bigl[\partial_{t}\delta_{ij}+\partial_{i}
\partial_{j}U\bigr]=exp\bigl\{{1\over
2}\int_{t_{1}}^{t_{2}}dt\partial^{2}U(x(t))\bigr\}\eqn\mangime$$
Hence the final result for the fermionic kernel is
$$\eqalign{{\bf K}_{F}(\psi_{2},t_{2} \ \vert  & \ \psi_{1},t_{1};[x])=
\cr \ = & \ {\bf\delta}\bigl(\psi_{2}-S(t_{2};[x])\psi_{1}\bigr)exp\bigl\{
{1\over 2}\int^{t_{2}}_{t_{1}}dt
\partial^{2}U\bigl(x(t)\bigr)\bigr\}\cr}\eqn\frumento$$
It is possible to check eq.\frumento~ also without referring to
path-integral manipulations. The path-integral on the RHS of eq.
\marton~ is the formal solution of the Schroedinger equation:
$$\bigl[{1\over 2}\partial^{2}_{i}U(x)-\partial_{i}\partial_{j}
U(x)\psi_{j}{\partial\over\partial\psi_{i}}+\partial_{t}\bigl]
{\bf K}_{F}(\psi,t\vert \psi_{1},t_{1};[x])=0\eqn\macina$$
with
~${\bf K}_{F}(\psi,t_{1}\vert\psi_{1},t_{1};[x])={\bf\delta}(\psi-\psi_{1})$.
It can be checked that the RHS of ~\frumento~ does indeed solve this
initial value problem (for ~$t>0$).
For future reference we note that ~${\bf K}_{F}$~can also be written as
$$\eqalign{{\bf K}_{F}(\psi_{2},t_{2}\ \vert & \ \psi_{1},t_{1};[x])=\cr
\ = & \ {\bf\delta}\bigl(S(t_{2};[x])^{-1}\psi_{2}-\psi_{1}\bigr)exp
\bigl\{-{1\over 2}\int^{t_{2}}_{t_{1}}dt
\partial^{2}U(x(t))\bigr\}\cr}\eqn\forca$$
because ~${\dot S}=-MS$~ implies that
$$det\bigl[S(t)\bigr]=exp\bigl\{-\int^{t}_{t_{1}}dt^{\prime}
\partial^{2}U(x(t^{\prime})\bigr\}\eqn\palo$$
Obviously the Hilbert space ~${\cal H}$~ of our theory is a direct
sum of subspaces with a fixed ghost number p, $0\le p \le N$.
Instead of labelling the various sectors by their ghost
number (or, equivalently, their degree as a differential form)
we shall use ~$f\equiv N-p$, ~$0\le f \le N$, so that the
subspace ~${\cal H}_f$~ of ~${\cal H}$~ consists of wave functions of the
form~\forse. Consequently
$${\cal H}=\bigoplus_{f=0}^{N}{\cal H}_{f}\eqn\sottana$$
The partition function decomposes correspondingly
$$\eqalign{Z_{susy}(T)\ \equiv & \ Tr\bigl[exp(-TH_{susy})\bigr]\cr
\ = & \ \sum_{f=0}^{N}Z_{f}(T)\cr}\eqn\pantaloni$$
with
$$Z_{f}(T)=\sum_{\alpha}\bra{\Phi^{\alpha}_{f}}exp(-TH_{f})\ket
{\Phi^{\alpha}_{f}}\eqn\calze$$
where ~$H_{f}$~ denotes the restriction of ~$H_{susy}$~to
~${\cal H}_{f}$, and ~$\bigl\{\Phi^{\alpha}_{f}\bigr\}$~is a
complete set of states  with ghost number p=N-f. The Hamiltonians
are particularly simple for f=0 and f=N. In these cases they are
diagonal in the tensor indices:
$$\eqalign{H_{0}\ = & \ -{1\over 2}\partial^{2}+{1\over 2}\partial
_{i}U\partial_{i}U-{1\over 2}\partial^{2}U\cr
H_{N} \ = & \ -{1\over 2}\partial^{2}+{1\over 2}\partial_{i}U
\partial_{i}U+{1\over 2}\partial^{2}U\cr}\eqn\scarpe$$
The respective partition functions can be represented by
purely bosonic path-integrals:
$$Z_{0,N}(T)=\int_{pbc}{\cal D}x(t)~exp \bigl\{-\int_{0}^{T}
dt~{\it L}_{0,N}\bigr\}\eqn\ciabatte$$
where
$$\eqalign{ {\it L}_{0} \ = & \ {1\over 2}{\dot x}_{i}^{2}+{1\over 2}
\partial_{i}U\partial_{i}U-{1\over 2}\partial^{2}U\cr
{\it L}_{N} \ = & \ {1\over 2}{\dot x}_{i}^{2}+{1\over 2}
\partial_{i}U\partial_{i}U+{1\over 2}\partial^{2}U\cr}\eqn\camicia$$
are the restriction of ~${\it L}_{susy}$ to ~${\cal H}_{0}$~
and ${\cal H}_{N}$, respectively. In ~\ciabatte~ we used periodic
boundary conditions (pbc): ~$x(0)=x(T)$. As a preparation
for the discussion of the Lyapunov exponents, we shall now give
an alternative representation of ~$Z_{0}$~ and ~$Z_{f}$~ in terms
of the complete supersymmetric lagrangian ~${\it L}_{susy}$ of
~\mediocre. For f=0 we may write
$$Z_{0}(T)=\int_{pbc}{\cal D}x(t)~\int {\cal D}{\psi}
(t){\cal D}{\bar\psi}(t)~exp\bigl\{-\int _{0}^{T}
dt~{\it L}_{susy}\bigr\}{\bf\delta}(\psi(0))\eqn\foulard$$
Again, periodic boundary conditions are used for the
~$x_{i}$-integration, but  $\psi(0)$ ~and ${\psi}(T)$~
are treated as independent integration variables. Note that these
boundary conditions are  different from the ones usually
used in supersymmetric quantum mechanics.
The equivalence of~\foulard~ and ~\ciabatte~ is established by noting
that:
$$\eqalign{\int{\cal D}\psi~\ {\cal D} & \ {\bar\psi}~exp\bigl\{
-\int_{0}^{T}{\bar \psi}_{i}\bigl[\partial_{t}\delta_{ij}+\partial_{i}
\partial_{j}U\bigr]\psi_{j}\bigr\}{\bf\delta}(\psi(0))=\cr
\ = & \ \int d\psi(T)~d\psi(0)~{\bf K}_{F}(\psi(T),T\vert \psi(0),0;[x] )
{\bf\delta}(\psi(0))\cr
\ = & \ exp\bigl\{{1\over 2}\int^{T}_{0}dt
\partial^{2}U(x(t))\bigr\}\cr}\eqn\maglione$$
where ~\frumento~ has been used in the last line. The term
~$\sim \partial^{2}U$~ found in ~\maglione~, when combined with the
first two terms on the RHS of eq.\mediocre, yields exactly the
lagrangian ~${\it L}_{0}$~of~\camicia. In the same way one can show
that
$$Z_{N}(T)=\int_{pbc}{\cal D}x(t)\int {\cal D}\psi(t) {\cal D}
{\bar\psi}(t)~{\bf\delta}(\psi(T))~exp\bigl\{-\int_{0}^{T}dt
{\it L}_{susy}\bigr\}\eqn\maglion$$
(Here and in the following we ignore multiplicative
constants.)
\par
In deriving eq.~\frumento~ we realized already that the dynamics of
the Grassmannian variables ~$\psi_{i}(t)$~ is essentially
deterministic. In fact, given a fixed trajectory ~$x_{i}(t)$, the kernel
~${\bf K}_{F}$~ is non-zero only if the initial and final value of ~$\psi$~
are related by the " classical" Jacobi matrix
~$S(t_{2};[x])$. Because all the information about Lyapunov
exponents is contained in ~$S$, this suggests that it should be
possible to extract them from certain correlation functions
involving ghosts and antighosts as we did in the deterministic
case. In fact, let us consider
$${\Gamma}_{f}(T)=\VEV{\psi_{k_{1}}(T)\psi_{k_{2}}(T)\cdots
\psi_{k_{f}}(T){\bar\psi}_{k_{f}}(0)\cdots{\bar\psi}_{k_{2}}(0)
{\bar\psi}_{k_{1}}(0)}\eqn\manica$$
with the expectation value~$\VEV{{\cal O}}$ of observables
~${\cal O}$~ defined by
$$\VEV{{\cal O}}\equiv Z_{0}(T)^{-1}\int_{pbc}
{\cal D}x\int {\cal D}\psi {\cal D}{\bar\psi}~{\cal O}
~{\bf\delta}(\psi(0))~exp\bigl\{-\int^{T}_{0} dt {\it L}_{susy}\bigr\}
\eqn\polsino$$
where ~$\psi(T)$~ and ~$\psi(0)$~ are independent integration
variables. As we discussed in section 4,
the operator ~${\bar\psi}_{k_{f}}\cdots{\bar\psi}_{k_{1}}$~
creates at time t=0 a f-volume from the vacuum which is
propagated by the supersymmetric dynamics until it is destroyed
by ~$\psi_{k_{1}}\cdots\psi_{k_{f}}$~ at time t=T. This becomes
obvious if we use ~\mediocre~ in ~\polsino, in order to write
$${\Gamma}_{f}(T)=Z_{0}^{-1}\int_{pbc}{\cal D}x~exp\bigl\{-\int_{0}^{T}
dt\bigl[{1\over 2}{\dot x}_{i}^{2}+{1\over 2}(\partial_{i}U)^{2}
\bigr]\bigr\}{\cal G}_{f}(T;[x])\eqn\calzino$$
with
$$\eqalign{{\cal G}_{f}(T;[x]) \ = & \ \int{\cal D}\psi~{\cal D}
{\bar\psi}~
exp\bigl\{-\int^{T}_{0}dt~{\bar\psi}_{i}\bigl[\partial_{t}
\delta_{ij}+\partial_{i}\partial_{j}U\bigr]\psi_{j}\bigr\}\cr
\ {\cdot} & \ \psi_{k_{1}}(T)\cdots\psi_{k_{f}}(T){\bar\psi}_{k_{f}}(0)
\cdots{\bar\psi}_{k_{1}}(0)~{\bf\delta}(\psi(0))\cr
\ = & \ \int d^{N}\psi(T)~d^{N}\psi(0)~\psi_{k_{1}}(T)\cdots
\psi_{k_{f}}(T)~{\bf K}_{F}(\psi(T),T\vert \psi(0),0;[x])\cr
\ {\cdot} & \ {\partial\over\partial\psi_{k_{f}}(0)}\cdots
{\partial\over\partial\psi_{k_{1}}(0)}{\bf\delta}(\psi(0))\cr}\eqn\lunga$$
Using ~\frumento~ it is easy to abtain:
$${\cal G}_{f}(T,[x])=exp\bigl\{{1\over 2}\int^{T}_{0}dt~\partial
^{2}U(x)\bigr\}\cdot S^{[k_{1}}_{k_{1}}S^{k_{2}}_{k_{2}}\cdots
S_{k_{f}}^{k_{f}]}\eqn\anticom$$
with ~$S^{i}_{j}\equiv S_{ij}(T,[x])$ ~(we do not distinguish
between upper and lower indices here). Inserting
~\anticom~into~\calzino, and making use of ~\camicia~,
we obtain the final result
$${\Gamma}_{f}(T)=\VEV{\VEV{S^{[k_{1}}_{k_{1}}S^{k_{2}}_{k_{2}}
\cdots S^{k_{f}]}_{k_{f}}}}\eqn\parentes$$
with the average~$\VEV{\VEV{\cdots}}$~ performed by means of the
lagrangian ~${\it L}_{0}$,~\ie,
$$\VEV{\VEV{{\cal O}[x]}}\equiv~Z_{0}(T)^{-1}\int_{pbc}
{\cal D}x~{\cal O}[x]~exp\bigl\{-\int^{T}_{0}dt{\it
L}_{0}\bigr\}\eqn\palla$$
Before relating~\parentes~ to the Lyapunov exponents, let us briefly
discuss another representation of ~$\Gamma_{f}(T)$. We shall show that
~${\Gamma_{f}}$~ is the ratio of the partition functions of
~${\it H}$ and ~${\it H}_{0}$, respectively:
$$\Gamma_{f}(T)={Z_{f}(T)\over Z_{0}(0)}\equiv
{Tr\bigl[exp(-TH_{f})\bigl]\over
Tr\bigl[exp(-TH_{0})\bigr]}\eqn\bastone$$
In order to evaluate ~$Z_{f}$~ as written down in eq.~\calze~, we
first note that the components of the
vector~$exp(-TH_{f})\Phi^{\alpha}_{f}$~ can be represented by the
bosonic functional integral
$$\bigl(exp(-TH_{f})\Phi^{\alpha}_{f}\bigr)(x,\psi)=\int {\cal D}
x~exp\bigl\{-\int_{0}^{T}dt{\it L}_{N}\bigr\}\Phi^{\alpha}
_{f}\bigl(x(0),S^{-1}(T,[x])\psi\bigr)\eqn\carota$$
with ~$x(T)=x$~ and where~${\cal D}x$~ includes an integration over
$x(0)$. To arrive at eq.~\carota~, we used ~\maria~ with~\marto~ and
~\forca, as well as the definition of ~${\it L}_{N}$~ in ~\camicia.
Furthermore, the inner product of two p-forms of the type~\belletti~
is given by
$$<\Phi^{(1)}\vert \Phi^{(2)}>={1\over p!}\int
d^{N}x~\Phi^{(1)\star}_{k_{1}\cdots k_{p}}(x)\Phi^{(2)}_{k_{1}\cdots
k_{p}}(x)\eqn\prodotto$$
If we use~\carota~with ~\prodotto~ in ~\calze, and exploit the
completness relation of the ~$\Phi^{\alpha}_{f}$,~we arrive at:
$$Z_{f}(T)=\int_{pbc}{\cal D}x~ exp\bigl\{-\int_{0}^{T}{\it
L}_{N}\bigr\}(S^{-1})^{[k_{1}}_{k_{1}}\cdots
(S^{-1})^{k_{N-f}]}_{k_{N-f}}\eqn\somma$$
The occurence of N-f factors of the Jacobi matrix is due to the
fact that ~${\cal H}_{f}$~ consists of moniomials with N-f
factors of~$\psi$,~\ie, differential forms of degree N-f. However,
if we use the identity
$$S^{[k_{1}}_{k_{1}}\cdots S^{k_{f}]}_{k_{f}}=C^{f}_{N}~det (S)
(S^{-1})^{[k_{1}}_{k_{1}}\cdots
(S^{-1})^{k_{N-f}]}_{k_{N-f}}\eqn\differenza$$
with ~$det(S)$ given by eq.~\palo, and $C^{f}_{N}$ some constants,
we see that ~$Z_{f}(T)$ coincides
with ~$Z_{0}(T)\cdot \Gamma_{f}(T)$ as given in eq.~\parentes~.
%\foot{The ~$C^{f}_{N}$ in ~\differenza~ are  overall constants.}.
This completes the proof of ~\bastone.
\par
Benzi et al.\refmark{8} and Graham\refmark{9} have studied eq.
\parentes~for the special case of a {\it one-dimensional} configuration
space, \ie, N=1. They determined a generalized Lyapunov expo            nent
~$\Lambda_{1}$~ from the behaviour of ~$\Gamma_{1}(T)$~ for
~$T\rightarrow\infty$~: ~$\Gamma_{1}(T)\sim exp(\Lambda_{1}T)$.
If we represent ~$\Gamma_{1}$~ as the ratio of two partition
functions, like in eq.~\bastone, we see that ~$\Lambda_{1}$~
can be expressed in terms of the lowest eigenvalues
~$E_{0}^{min}$~ and ~$E_{1}^{min}$~ of ~$H_{0}$~ and ~$H_{1}$~,
respectively:
$$\Lambda_{1}=E^{min}_{0}-E^{min}_{1}\eqn\divisione$$
If supersymmetry is unbroken, the vacuum is non-degenerate
so that \break either ~$E^{min}_{0}< E^{min}_{1}$ ~or ~$E^{min}_{0}>
E^{min}_{1}$. Only in the second case, when the vacuum is in
the one-form sector, a positive Lyapunov exponent is obtained.
\par
In the work of Benzi et al.\refmark{8} and Graham\refmark{9}
only one dimensional systems were considered, so that only the
exponent~$\Lambda_{1}$~ related to the evolution of
one-dimensional volume elements could be defined. In the
present paper we suggest the following interpretation of the
correlation function ~$\Gamma_{f}(T)$~ of~\parentes.
The antisymmetrized product of ~$S$-matrices describes the
evolution of f-dimensional volume elements. Therefore the
large-T behaviour of ~$\Gamma_{f}(T)$~ can be used to define
higher dimensional Lyapunov-like exponents:~$\Lambda_{f}$,~
$1\le f\le N$. In view of eq.(2.10) and the discussion in section 4,
it is natural to define
$$\Gamma_{f}(T)\sim exp\bigl[(\Lambda_{1}+\Lambda_{2}+\cdots
\Lambda_{f})T\bigr]\eqn\cavata$$
for ~$T\rightarrow\infty$. This is the higher
dimensional generalization of eq. (2.11) discussed in the
introduction. By virtue of eq.~\bastone~ the sum of the
first f ~$\Lambda$'s is given by the lowest eigenvalue ~$E^{min}_{f}$
in the f-form sector:
$$\Lambda_{1}+\Lambda_{2}+\cdots
+\Lambda_{f}=E_{0}^{min}-E^{min}_{f}\eqn\tappo$$
The above relation was conjectured in ref.[9] but no proof was provided,
while here, using the machinery developed above, we have given a proof of
it. For the deterministic case we cannot have a relation like
~\tappo~ because the spectrum of the relative Hamiltonian~${\widetilde{\cal
H}_{f}}$~is not bounded below
\endpage
\Appendix{B}
In this appendix we briefly comment on the relation between our 
path-integral formalism and the thermodynamic one of Ruelle\refmark{23}.
The basic object of ref.[23] is a Fredholm determinant
that is by now  called the Ruelle Zeta-function
~$\zeta(z)$~is defined as 
$$\zeta(z)=det\bigl[1-z{\bf K}_{(0)}\bigr]= exp~\bigl[-\sum_{n=1}^{\infty}
{z^{n}\over n}~Tr({\bf K}_{(0)}^{n})\bigr]\eqn\biuno$$
where ~$z$~is a complex variable and ~${\bf K}_{0}$~ (for the classical
Hamiltonian systems) is the kernel
represented in eq.~\ford~with the ghosts ~$c^{a}$~put to zero and
the interval of time taken to be a finite one which we will call
 ~$\Delta$. So, roughly speaking, ~$K(\Delta)$~ are the matrix
elements of ~$exp~[-i{\widetilde{\cal H}}_{0}\Delta]$~which is the
operator of evolution for zero-forms.
The above series  can be defined for a wide class of maps and it has
a finite radius of convergence. The operator 
$exp~[-i{\widetilde{\cal H}}_{0}\Delta]$~ is also called the Koopman
operator and its inverse the Perron-Frobenius operator\Ref\Lasot{A.Lasota
 and M.MacKey, "{\it Probabilistic properties of deterministic
systems}",\nextline
Cambridge University Press, 1985, Cambridge UK}
Using the formalism of Ruelle, it is easy to prove that the series ~\biuno~ 
is also equal to the following one
$$\zeta(z)=exp\bigl[-\sum_{n=1}^{\infty}{z^{n}\over n}a_{n}\bigr]\eqn\bidue$$
where, (for the Hamiltonian evolution of zero-forms),
$$a_{n}= number~of~periodic~trajectories~of~period~n\Delta\eqn\bitre$$
Note that, using eqn.~\pistoia, we can say that
$$a_{n}=Z_{o}(n\Delta)\eqn\biquattro$$
Here~$Z_{0}$~has to be evaluated with all the cares we indicated
under equations ~\pistoia,~ basically following the lines of
the last of refs.[23].
Ruelle proved the relation ~\bidue~ for very general maps
and to do that he had to use a highly sophisticated mathematical
machinery. At a formal level, for classical hamiltonian systems,
the proof is straightforward  if we use our path-integral representation 
\refmark{10}. In fact, following the steps ~\fraschini~through~\ford~
we see that the trace of the Koopman operator boils down to  an integral 
over a Dirac delta function
$$\Tr\bigl[exp\bigl(-i{\widetilde{\cal
H}}_{0}(n\Delta)\bigr)\bigr]=Z_{o}(n\Delta)$$
and this proves the theorem above. We do not pretend any mathematical
rigorosity in the proof we have given above, but we have provided it
to give to the reader some intuition about the Ruelle Zeta-function.
Another point to be cautious in the above proof is the problem of the
continuum limit. The Ruelle zeta function was always built for discrete
maps and {\it not} for the continuous time evolution given by our
path-integral. So in the proof above,~$Z_{0}(n\Delta)$~should be
considered  the finite-lattice approximation to our path-integral,
and actually that is how path-integrals are defined. Taking the limit
of ~$\Delta$~going to zero has to be done with the same care as explained
in the book of Feynman-Hibbs\Ref\Feyn{R.P.Feynman and A.R.Hibbs, "{\it
Quantum Mechanics and path-integrals}",\nextline
McGraw-Hill, 1965, New York} for standard path-integral.
If we want to avoid that, then we could choose ~$\Delta$~as long as
the whole interval of time we are interested. Doing that we can then use the
Ruelle Zeta-function to derive our ~$Z_{0}(\Delta)$, in fact
it is easy to see that
$$Z_{0}(\Delta)=Tr\bigl[ exp~\bigl(-i({\widetilde{\cal H}}_{0}\Delta)\bigr)
\bigr]=-\oint {dz\over (2\pi i)}{ln\zeta(z)\over z^{2}}\eqn\bisei$$
So one see that, using the appropriate zeta-function, we can extract
one of the partition functions we have used in the paper.
\par
The same construction  can also be applied to the
kernel of evolution of higher forms and this was also already
envisioned by Ruelle in 1976\refmark{23}.
We can define a {\it generalized} Ruelle zeta-function as
$$\zeta_{(p)}(z)=det\bigl[1-z{\bf K}_{(p)}\bigr]=exp~\bigl[-\sum_{
n=1}^{\infty}{z^{n}\over n}Tr({\bf K}_{(p)}^{n}\bigr]\eqn\bisette$$
where we called ~${\bf K}_{(p)}$~the operator~$exp~-i{\widetilde
{\cal H}}_{p}\Delta$~appearing in~\parma~which makes the evolution
of p-forms. Like for the zero-form case,
for these generalized zeta functions~$\zeta_{(p)}(z)$ 
there exists an alternative representation, namely
$$\zeta_{(p)}(z)=exp\bigl[-\sum {z^{n}\over n}a_{n}^{(p)}\bigr]\eqn\biotto$$
where the coefficients ~$a_{n}^{(p)}$~are given by
$$a_{n}^{(p)}=\sum_{\phi_{per}}\bigl[S^{-1}(n\Delta;\phi_{per})^{[a_{1}}_{a_{1}}
\cdots S^{-1}(n\Delta;\phi_{per})^{a_{p}]}_{a_{p}}\bigr]\eqn\binove$$
Here ~$S^{a}_{b}$~is the Jacobi matrix we introduced in 
eq.~\benevento~and~$\phi_{per}$~ denotes all points from which
a periodic orbit of period ~$n\Delta$~originates.
From eq.~\binove~and~\livorno~we get immediately that
$$a_{n}^{(p)}=Z_{p}(n\Delta)\eqn\bidieci$$
The proof of the above theorem is basically what is contained 
in formulas~\parma~through~\livorno.\break One could have also used directly the
~${\bf K}$~of eq.~\ford~
and have all the higher forms included in a single formula:
$${\tilde\zeta}(z)=det \bigl[1-z{\bf K}\bigr]=exp[-\sum_{n=1}^{
\infty}Tr({\bf K}^{n})\bigr]\eqn\biundici$$
The trace ~$Tr{\bf K}^{n}$~ can be taken either
by choosing periodic boundary conditions or antiperiodic ones for the ghosts. 
In the second case, which corresponds to Ruelle's choice\refmark{23}, 
we will get an expression which is an alternating
sum (according to the order of the forms) and which
will not depend at all on the Hamiltonian.  This has been proved in
the third of refs.[10] along the lines of  modern
topological field theory\Ref\wito{E.Witten, Com.Math.Phys.117 (1988) 353;
ibid. 118 (1988) 130;\nextline
D.Birmingham et al.,~ Phys.Rep.209 (1991) 130}. It is remarkablethat
the same trick of getting something which does not depend on
the Hamiltonian, by working with an alternating sum, was already
employed by Ruelle in 1976.
In our paper on the contrary, we have exploited also the
individual terms of this alternating sum, and seen that they are  related to
higher order Lyapunov exponents.
Finally we  mention also the work of Christiansen et al.\Ref\Pal{
F.Christiansen et al., J.Phys.A Math.Gen. 23 (1990) L1301} where 
forms were used in order to get information on the dynamics
and not just on topological features of the manifold. In this
work\refmark{29} forms have a different meaning than
in our formalism, however.
\ack
We wish to thank A.Comtet  who brought to our
attention refs.[8] and G.Parisi for referring us to the
work of Ruelle\refmark{23}.
M.R. wishes to acknowledge the hospitality of the
department of theoretical physics of the University of Trieste.
This work has been  supported in part by grants brom  INFN, MURST 
and NATO.

\refout
\vfil\eject
\bye